\documentclass[12pt,a4paper]{article}
\pdfoutput=1

\usepackage[T1]{fontenc} 
\usepackage{amsmath}
\usepackage{amsfonts}
\usepackage{amssymb}
\usepackage{graphicx}
\usepackage{subfig}
\usepackage{hyperref}
\usepackage{color}
\usepackage{verbatim}
\usepackage[makeroom]{cancel}
\usepackage{color}
\usepackage{cite}
\newcommand{\beq}{\begin{equation}}
\newcommand{\eeq}{\end{equation}}
\newcommand{\bea}{\begin{eqnarray}}
\newcommand{\nn}{\nonumber}
\newcommand{\eea}{\end{eqnarray}}
\def\be{\begin{equation}}
\def\ee{\end{equation}}

\def\beq{\begin{equation}}
\def\eeq{\end{equation}}

\def\({\left(}
\def\){\right)}

\def\mpl{M_{\rm Pl}}
\def\p{\partial}

\def\lsim{\mathrel{\rlap{\lower3pt\hbox{\hskip0pt$\sim$}}
     \raise1pt\hbox{$<$}}}         
\def\gsim{\mathrel{\rlap{\lower4pt\hbox{\hskip1pt$\sim$}}
     \raise1pt\hbox{$>$}}}         
\def\lsim{\mathrel{\rlap{\lower3pt\hbox{\hskip0pt$\sim$}}
     \raise1pt\hbox{$<$}}}         
\def\gsim{\mathrel{\rlap{\lower4pt\hbox{\hskip1pt$\sim$}}
     \raise1pt\hbox{$>$}}}         
\def\beq{\begin{eqnarray}}
\def\eeq{\end{eqnarray}}
\def\ba{\begin{eqnarray}}
\def\ea{\end{eqnarray}}
\def\({\left(}
\def\){\right)}

\def \be {\begin{equation}}
\def \ee {\end{equation}}
\def \beal {\begin{equation}\begin{aligned}}
\def \eeal {\end{aligned}\end{equation}}
\def \bealsn {\begin{equation*}\begin{aligned}}
\def \eealsn {\end{aligned}\end{equation*}}
\def \besn {\begin{equation*}}
\def \eesn {\end{equation*}}
\def \bea {\begin{eqnarray*}}
\def \eea {\end{eqnarray*}}
\def \bear {\begin{eqnarray}}
\def \eear {\end{eqnarray}}
\def \bn {\begin{enumerate}}
\def \en {\end{enumerate}}
\def \bi {\begin{itemize}}
\def \ei {\end{itemize}}
\def \bex {\begin{Exercise}}
\def \eex {\end{Exercise}}
\def \ban {\begin{Answer}}
\def \ean {\end{Answer}}
\def \bde {\begin{description}}
\def \ede {\end{description}}
\def \dd  {{\rm d}}

\begin{document}
\def\thefootnote{\fnsymbol{footnote}}

\begin{center}
\Huge{{\bf Gaugid inflation }}\\

\end{center}
\begin{center}
\vspace{.3cm}

\vspace{.5cm}


\large{Federico Piazza$^{\rm a}$, David Pirtskhalava$^{\rm b}$, Riccardo Rattazzi$^{\rm b}$, Olivier Simon$^{\rm b}$}
\\[0.7cm]

\normalsize{
\textit{$^a$ CPT, Aix-Marseille University, Universit\'e de Toulon, CNRS UMR 7332 \\
13288 Marseille, France}}
\vspace{.2cm}

\normalsize{
\textit{$^b$Theoretical Particle Physics Laboratory, Institute of Physics\\
EPFL, CH-1015 Lausanne, Switzerland}}
\vspace{.2cm}

\vspace{.8cm}

\hrule \vspace{0.3cm}
\noindent \small{\textbf{Abstract}\\
} 
\end{center}
The spectrum of  primordial gravitational waves is one of the most robust inflationary observables, often thought of as a direct probe of the energy scale of inflation.~We present a simple model, where the dynamics controlling this observable is very different than in the standard paradigm of inflation.  The model is based on a peculiar finite density phase---the \textit{magnetic gaugid}---which stems from a highly non-linear effective theory of a triplet of abelian gauge fields. The gaugid extends the notion of homogeneous isotropic solid, in that its spectrum of fluctuations includes helicity-2 phonons. We show how, upon implementing the gaugid to drive inflation, the helicity-2 phonon mixes with the graviton, significantly affecting the size of the primordial tensor spectrum. The rest of the features of the theory, in particular the vector and scalar perturbations, closely resemble those of \textit{solid inflation}. 
\\ 
\noindent
\hrule
\def\thefootnote{\arabic{footnote}}
\setcounter{footnote}{0}

\newpage

\section{Introduction}

Little is known about the fundamental physics behind inflation, and indeed a large number of models can account for the existing observations reasonably well. With no single `best-motivated' proposal at hand, it proves useful to resort to a maximally model-independent, effective field theory (EFT)-based treatment of inflation. Such EFT approach is a concrete consequence of the reckoning that {\it {cosmology is nothing but gravity at play  with condensed matter}}: FRW solutions to Einstein equations and their small perturbations are indeed propelled by an energy momentum tensor $T_{\mu\nu}$ describing a homogeneous and isotropic finite density phase of matter (with  ``matter'' here broadly defining any degree of freedom other than gravity). In this situation the long wavelength perturbations of the matter sector, the hydrodynamic modes, are nothing but the Golstone bosons associated to the spontaneous breaking of spacetime and internal symmetries down to the residual group $ISO(3)$ 
characterizing a homogeneous and isotropic medium \cite{Nicolis:2015sra}. 
Very much like it happens for the pions  of QCD, the dynamics of these Goldstone bosons  can be systematically described by an effective lagrangian that is purely determined by the pattern of symmetry breaking. In particular,  
 the details of the microphysics are efficiently and systematically encoded in a limited set of ``relevant'' parameters. As gravity just gauges the Poincar\'e group, its interplay with the Goldstones is similarly dictated by symmetry, hence the great power of the EFT approach to the early universe. 

\vskip 0.15cm 
 
As illustrustrated in Ref.~\cite{Nicolis:2015sra}, there exist various  patterns of
symmetry breaking realizing  $ISO(3)$ as residual symmetry, each characterized by its own Goldstone boson spectrum and dynamics. There exists then in principle a corresponding variety of early cosmologies.  Nonetheless, the vast majority of inflationary models considered so far are based on scalar fields
and on the patterns of symmetry breaking they can realize. 
Single field inflation offers  the simplest option to break the Poincar\'e group $ISO(3,1)$ down to $ISO(3)$. Indeed, a single scalar with a time-dependent expectation value breaks time translations and  Lorentz boosts in one shot. Schematically, 
\beal
\label{tt0}
\phi(t) \ \Longrightarrow \ \{\cancel{K_i},~ \cancel{P_0}\} ~.
\eeal
In the presence of gravity, the appropriate gauged version of this statement  is that time-diffeomorphisms are spontaneously broken (nonlinearly realized) while reparametrizations of the uniform-time hypersufaces remain intact:
\begin{equation}
\label{tt}
  \text{diffs} =  \left \{
  \begin{aligned}
    & t\to t+\xi^0(t,\vec{x}) &&\text{broken} \\
    &x^i\to x^i+\xi^i(t,\vec{x}) &&\text{unbroken}~.
  \end{aligned} \right. ~
\end{equation} 
Most of the features of single field inflation directly follow from this symmetry breaking pattern. 
In particular, the physics of the corresponding Goldstone boson (also known as the \textit{adiabatic mode}) is governed by a robust structure \cite{Creminelli:2006xe,Cheung:2007st} that entails conservation of the physical curvature perturbation on super-horizon scales and a set of ``consistency relations'' between different $n$-point functions that stem from the associated Ward identities~\cite{Maldacena:2002vr,Creminelli:2004yq,Cheung:2007sv,Hinterbichler:2012nm,Creminelli:2012qr,Hinterbichler:2013dpa,Bordin:2017ozj}.
Needless to say, inflationary theories characterized by the breaking pattern \eqref{tt} have been explored extensively over the years -- both on model-by-model basis as well as within the more model-independent, effective field theory framework \cite{Senatore:2009gt,Senatore:2010wk,Piazza:2013coa,Pirtskhalava:2015ebk}. 

\vskip 0.15cm

The symmetry breaking pattern \eqref{tt} is rather constraining, yet it does allow for some freedom---especially when it comes to the scalar perturbations. Indeed, the primordial scalar spectrum is quite sensitive to the precise details of the underlying theory: it can be modified at will by changing the shape of the potential, the  propagation  speed of the perturbations~\cite{Garriga:1999vw,Alishahiha:2004eh} and/or their long-wavelength dispersion relations~\cite{ArkaniHamed:2003uz}, or by introducing extra fields~\cite{Linde:1996gt,Enqvist:2001zp,Lyth:2001nq,Dvali:2003em}.~In contrast, the tensor perturbations are  far more robust in that their properties are unambiguously determined by the breaking pattern \eqref{tt}.~In particular,  the unbroken spatial diffeomorphisms forbid  a small ``mass'' term for the graviton, which would affect the tilt of the tensor spectrum. Likewise, the amplitude of the latter observable is hardly amenable to modification (the first correction comes from a sub-leading parity-violating EFT operator with three derivatives  \cite{Creminelli:2014wna}) and is given by
\beal
\label{stdspec}
\Delta^2_t\sim \frac{H^2}{\mpl^2}~. 
\eeal

\vskip 0.15cm

Beyond eqs.~\eqref{tt0} and \eqref{tt}, there exist however other inequivalent patterns of symmetry breaking preserving homogeneity and isotropy \cite{Nicolis:2015sra}. These possibilities include systems whose microscopic dynamics breaks purely spatial symmetries.
One example of such a system is provided by a \textit{solid}, which at distances larger than its crystalline structure can be modelled by a triplet of scalar fields  $\phi^I$ parametrizing the comoving coordinates of its elements \cite{Dubovsky:2005xd}. In addition to breaking Lorentz boosts, the solid's ground state 
\be
\label{solidbckd}
\langle \phi^I\rangle = x^I
\ee
spontaneously breaks the three spatial translations and three rotations---i.e. the entire space-like subgroup $ISO(3)_{P}$ of the Poincar\'e group. Such a configuration is nevertheless consistent with homogeneity and isotropy provided that the underlying Lagrangian is invariant under the group $ISO(3)_{I}$ of \textit{internal} translations and rotations, acting on the scalar triplet.~While $ISO(3)_{I}$ is fully broken by the ground state as well, the diagonal subgroup of $ISO(3)_{P}\times ISO(3)_{I}$ remains unbroken and it is precisly the symmetry under the latter group---referred to here as $ISO(3)_{\text{diag}}$---that ensures homogeneity and isotropy of the long wavelength dynamics.~Upon coupling to gravity, the system provides alternative means of driving cosmic acceleration \cite{Bucher:1998mh,Gruzinov:2004ty,Endlich:2012pz}.  

\vskip 0.15cm

All the distinctive properties of \textit{solid inflation} \cite{Endlich:2012pz} arise because eq.~\eqref{solidbckd} 
realizes another  symmetry breaking pattern than \eqref{tt}. This results in a peculiar phenomenology, not captured by the standard EFT of inflation of refs.~\cite{Creminelli:2006xe,Cheung:2007st}. For example, the spectrum of the primordial density perturbations acquires a unusual scaling $\Delta^2_s \propto c_s^{-5}$ with respect to the scalar speed of sound $c_s$~.~Furthermore, the underlying symmetry breaking pattern does allow  the graviton to acquire a small mass, $m_\gamma^2\sim -\dot H$, leading to a slightly blue-tilted tensor spectrum, uncharacteristic of conventional inflationary theories. 

\vskip 0.15cm

As to the \textit{amplitude} of the tensor perturbations, it does retain the standard form \eqref{stdspec} in solid inflation, putting robustness of the primordial tensor spectrum on even firmer grounds.
As a matter of fact, the existing limits on the primordial tensor-to-scalar ratio\footnote{The current upper bound is around $r\lsim 10^{-1}$ \cite{Array:2015xqh}, and is expected to get improved by two orders of magnitude in the foreseeable future.} are already highly constraining for a generic EFT of single-field inflation (see \cite{Pirtskhalava:2015ebk} for a recent analysis), so it is important to understand under what circumstances---if at all---it is possible to have predictions for the tensor spectrum that differ from eq.~\eqref{stdspec}.

\vskip 0.15cm

In this paper we wish to put forward a model of inflation whose dynamics lead to a primordial tensor spectrum that drastically differs from the standard result in eq.~\eqref{stdspec}. The (unconventional) order parameter for spacetime symmetry breaking is provided by a triplet of \textit{$U(1)$} gauge fields $A^I_\mu$ with a global internal $SO(3)_I$ acting on the $I$ index. The  internal symmetry group is similar to that of a solid, except that the three (internal) translations are part of the gauge group.  
Under broad conditions, the theory allows for a purely ``magnetic" vacuum
\be
\label{op}
\langle A^I_{\mu} \rangle = \varepsilon_{Ijk} \delta^j_
\mu x^{k}~,
\ee
that describes three mutually orthogonal homogeneous magnetic fields of equal norm
 \be
\label{mag}
\langle B^I_{j} \rangle =\delta^I_j~,
\ee
a configuration that clearly preserves a residual $ISO(3)_{\text{diag}}$.
  There is no counterpart to the background \eqref{op} among the possible ground states of non-Abelian gauge theories (with or without extra scalar fields), such as those proposed in~\cite{Maleknejad:2011jw,Adshead:2012kp} in the context of inflation.
We will refer to the above system as the \textit{magnetic gaugid} \cite{Nicolis:2015sra}. A related configuration is the \textit{electric gaugid}, characterized by the order parameter $\langle A^I_{\mu} \rangle \propto \delta^0_\mu x^I$. The electric gaugid, as will be clarified in Sec. 2, entails a slightly less minimal set up, and that is why we shall focus on the magnetic gaugid, which is a natural solution provided the original theory preserves parity $P$. 
\vskip 0.15cm
The spectrum of gaugid's excitations features more degrees of freedom than a solid does; in particular, it includes a pair of helicity-0 modes (scalars), an helicity-1 mode (vector), and an helicity-2 mode (tensor).
This makes up a total of 6 modes (helicity-1 and -2 come in both signs), corresponding to the underlying dynamics originating from three independent 
abelian gauge fields.~ The two scalars are respectively parity even and parity odd. Thus they do not mix at the linearized level. Curiously, the parity even scalar and vector modes are fully analogous to the scalar and vector perturbations in solid inflation \cite{Endlich:2012pz}. 

\vskip 0.15cm

The  helicity-2 phonon, which we refer to here as $E_{ij}$, is the remarkable novelty.~As we will show below, upon coupling the system to gravity, this mode mixes with the graviton via a single-derivative Chern-Simons-like operator
\beq
\mathcal{L} \sim  \varepsilon_{ijk} \p_k E_{il}\gamma_{lj}~.
\eeq
This significantly affects the properties of the primordial tensor spectrum.
One of the results of this mixing is that the tensor-to-scalar ratio in gaugid inflation has schematically the form
\beq
r\propto \frac{c^5_T}{c^3_E}\times (\epsilon_\gamma N_e)^2~,
\eeq
where $N_e$ is the number of    ``observable'' e-foldings, $\epsilon_\gamma\lsim 1/N_e$ is a slow roll parameter,  while $c_T$ and $c_E$ are the propagation velocities of the gaugid's parity-even scalar and tensor excitations respectively\footnote{For $\epsilon_\gamma$ somewhat larger than $N_e^{-1}$, the tensor-to-scalar ratio is saturated at $r\sim c_T^5/c_E^3$.} (both of these can be strongly subluminal, depending on the parameters of the theory). This is very different from the predictions for $r$ in both the conventional models of inflation governed by spontaneously broken time translations \eqref{tt}, as well as more exotic theories such as solid inflation. Measuring the primordial tensor spectrum is no longer synonymous with measuring the inflationary energy scale in the model under consideration, and sizeable gravitational waves can be generated even for a relatively low-scale inflation. In particular, for $\epsilon_\gamma\sim 1/N_e$ and $c_T\sim c_E$ one can even have $r\sim 1$, which is of course ruled out. On the reverse side, $r$ is very sensitive to the propagation velocity of the gaugid's scalar phonon, and becomes strongly suppressed for a somewhat subluminal $c_T$.
As to the tilt of the tensor spectrum, it can have both signs and is of the standard magnitude in most of the parameter space 
\be
|n_t|\sim  N_e^{-1}~.
\ee

\vskip 0.15cm

The paper is organized as follows.  In section \ref{gaugids} we characterize our effective field theory and study the conditions for  \eqref{op} to be a solution. Moreover we study the stress tensor corresponding to the latter configuration and derive the conditions under which it sources a quasi-de Sitter space. In section \ref{perts} we characterize the perturbations with and without gravity. In particular we study various gauge fixings and a residual symmetry that plays an important role in the study of perturbations. In Sec.~\ref{section4} we present a simple explicit 
model, free of instabilities and superluminal propagation, and study in detail the dynamics of its perturbations. In particular, we derive the expressions for the scalar and the tensor spectra. Sec.~\ref{reheating} discusses the epoch of reheating, for which, like in solid inlfation, extra assumptions are needed. We finally discuss a number of directions in which our work can be extended and conclude in Sec. \ref{conclusions}. 
\section{Gaugid cosmology}
\label{gaugids}

The order parameter for our symmetry breaking pattern is a triplet of $U(1)$ gauge fields $A_\mu^I$. In addition to the gauge redundancy, we will assume that the theory is invariant under internal  $SO(3)$ rotations, acting on the flavor indices. Therefore, the most general action for the system exhibits invariance under
\begin{equation}
A_\mu^I \ \longrightarrow \ A_\mu^I + \partial_\mu \chi^I, \qquad A_\mu^I \ \longrightarrow \ R_J^I A_\mu^J,
\end{equation}
where $\chi^I$ are the three gauge parameters (arbitrary functions of spacetime coordinates) and $R_J^I$ a three-dimensional constant rotation matrix. 
\subsection{The effective Lagrangian}
The effective Lagrangian is a function of all possible Lorentz- and $SO(3)_{ I}$ - invariant operators constructed out of the field strengths $F^I_{\mu\nu} = \p_\mu A^I_\nu - \p_\nu A^I_\mu~$ and their derivatives. In practice we will be interested in the physical situation where non-linearities in $F^I_{\mu\nu}$ are dynamically relevant, while the derivative expansion is convergent. That means that terms involving further derivatives on $F^I_{\mu\nu}$ are subdominant . A simple self-consistent instance where this state of things can be realized is broadly characterized by two parameters: a physical cut-off mass scale $\Lambda$, controlling both the non-linearities in the field and the derivative expansion, and a coupling $g_*$ describing the strength of the interaction at the scale $\Lambda$. Under these assumptions the lagrangian has the form
\beq
{\cal L}=\frac{\Lambda^4}{g_*^2} {\cal F}(F_{\mu\nu}^I/\Lambda^2,\partial_\mu/\Lambda)
\label{nda}
\eeq
in an obvious notation. By construction, $g_*$ controls the loop expansion at the scale $\Lambda$, and, 
as long as $g_*\lsim 4\pi$, the ansatz is  stable at the quantum level\footnote{This structure is ubiquitous in effective field theories derived from large N or string theory. Notice that $n$-point scattering amplitudes are $\propto g_*^{n-2}$.}.
Moreover, smooth field configurations with $|F^I_{\mu\nu}|\sim \Lambda^2$ and $|\partial_\mu| \ll \Lambda$ can in principle be well described within EFT, that is without exciting cut-off states, provided all the non linearities in $F^I_{\mu\nu}$ are kept. For the purpose of our study the scale $\Lambda$ will never directly appear.
It will thus be convenient to rescale $A^I_\mu\to A^I_\mu/\Lambda^2$, and absorb $\Lambda^4$ in eq.~(\ref{nda}) in the definition of ${\cal F}$. The rescaled $A^I_\mu$ carries dimension of length, consistently with its assumed background profile in eq.~(\ref{op}). 
%

\vskip 0.15cm
At the leading derivative level the Lagrangian is a function of $SO(3)\times SO(3,1)$ invariants built out of $F^I_{\mu\nu}$. Examples are
$F^I_{\mu\nu}F^{I\mu\nu}$, $F^I_{\mu\nu}F^{J\mu\nu} F^I_{\alpha\beta}F^{J\alpha\beta}$, etc..
In order to count the number of \textit{independent} such invariants, let us concentrate on Lorentz contractions first, \emph{i.e.}, on building Lorentz scalars with free $I, J,\dots$ indices.  To this end, we note that the field strength  tensor tranforms
under  the Lorentz group $SO(3,1) = SU(2)_L\times SU(2)_R$ as the direct sum $\bf (1, 0)\oplus \bf (0, 1)$. More precisely the chiral field strength
\be
\mathcal{F}^I_{\mu\nu} \ = \ F^I_{\mu\nu} +  \frac{i}{2}\varepsilon_{\mu\nu}^{~~\rho\sigma} F^I_{\rho\sigma} \ \equiv \ F^I_{\mu\nu} + i \tilde F^I_{\mu\nu} 
\ee
 transforms as $\bf (1, 0)$ and its complex conjugate $\bar{\mathcal{F}}^I_{\mu\nu}=F^I_{\mu\nu} - i \tilde F^I_{\mu\nu}$ as a $\bf (0, 1)$. In fact by using 't Hooft simbols $\eta^{\mu\nu}_a$ and ${\bar \eta}^{\mu\nu}_{\bar a}$, or directly from the above equation, the two irreps are written in terms of the electric and magnetic field as
 \beq
 ({\bf{1,0}})\equiv V^I_a=E^I_a+iB^I_a\qquad ({\bf{0,1}})\equiv \bar V^I_{\bar a}=E^I_{\bar a}-iB^I_{\bar a}
 \eeq
 with $\bar V^I_{\bar a}=(V^I_a)^*$.
 At this point, Lorentz scalars constructed out of the field strengths are simply given by (the real and imaginary parts of) the holomorphic $SU(2)_L$ invariants built out of $V^I_a$.
 No additional invariants involving both $V^I_a$ and $\bar V^I_{\bar a}$ can be built, because of their chiral tranformation properties.
There are only two such invariant contractions, namely, the (symmetric in flavor indices) bilinear $V^I_a V^J_a\equiv -\frac12(Y^{IJ}+i{\tilde Y^{IJ}})$ and the (totally antisymmetric) trilinear $\varepsilon^{abc}V^I_a V^J_b V^K_c\equiv U^{IJK}+i\tilde U^{IJK}$. The real and imaginary parts  are respectively P-even and P-odd and are easily found to be 
\begin{align}
Y^{IJ}\ \ &=  2 (\vec B^I \cdot \vec B^J - \vec E^I \cdot \vec E^J), \label{invariants}\\
\tilde Y^{IJ} \ \ &= - 2 (\vec B^I \cdot \vec E^J + \vec E^I \cdot \vec B^J),\\
U^{IJK}  &=  \vec B^I \cdot(\vec E^J \wedge \vec E^K) + \vec B^J \cdot(\vec E^K \wedge \vec E^I) +\vec B^K \cdot(\vec E^I \wedge \vec E^J) -  \vec B^I \cdot(\vec B^J \wedge \vec B^K) , \\
\tilde U^{IJK}  &=  - \vec E^I \cdot(\vec B^J \wedge \vec B^K) - \vec E^J \cdot(\vec B^K \wedge \vec B^I) -\vec E^K \cdot(\vec B^I \wedge \vec B^J) +  \vec E^I \cdot(\vec E^J \wedge \vec E^K) .
\label{realandim}\end{align}
The above 
 4 real invariants can also be written directly in terms of the $4\times 4$ field strengths as
\begin{equation}
Y^{IJ} =   F^{I}_{\mu\nu} F^{J\mu\nu}, \quad \tilde Y^{IJ}=  F^{I}_{\mu\nu} \tilde F^{J\mu\nu} , \quad U^{IJK} = F^{I \, \nu }_\mu F^{J \, \sigma}_\nu F^{K \, \mu }_\sigma, \quad 
\tilde U^{IJK} = F^{I \, \nu }_\mu F^{J  \, \sigma}_\nu \tilde F^{K \, \mu }_\sigma\, .
\end{equation}
In what follows we shall work in this spacetime index notation, as it is easier to incorporate the effects of the spacetime metric.
\vskip 0.15cm

We can now proceed to write down $SO(3)_I$-invariant combinations with the above ingredients. It is not difficult to see that there are a total of 11 invariants. Indeed, with the two matrices $Y^{IJ}$ and $\tilde Y^{IJ}$ we can form nine of these. One way to see this counting is to consider the two matrices $Y^{IJ}$ and $\tilde Y^{IJ}$  as described by the three eigenvalues of $Y^{IJ}$, the three eigenvalues of $\tilde Y^{IJ}$ plus the three parameters of the $SO(3)_I$ rotation that relate the two eigenbases. The trilinears, on the other hand, being totally antisymmetric, are already $SO(3)_I$ scalars: $U^{IJK} = U\varepsilon^{IJK}$, $\tilde U^{IJK} = \tilde U\varepsilon^{IJK}$. 

\vskip 0.15cm

To summarize, we can define the following set of independent internal $SO(3)$ and Poincar\'e invariants:
\be
\begin{split}
X &= F^{I}_{\mu\nu} F^{I\mu\nu} = [Y];\\[2mm]
I_1 &= \frac{[\tilde Y]}{[Y]}, I_2 = \frac{[Y^2]}{[Y]^2},  I_3 = \frac{[\tilde Y^2]}{[Y]^2},  I_4 = \frac{[Y \tilde Y]}{[Y]^2}, \  I_5 = \frac{[Y^3]}{[Y]^3}, \    I_6 = \frac{[Y^2 \tilde Y]}{[Y]^3},  I_7 = \frac{[\tilde Y^3]}{[Y]^3}, \\[2mm]
 I_8 &= \frac{[Y^3 \tilde Y]}{[Y]^4},  I_9 = \frac{U^{IJK}\varepsilon_{IJK}}{[Y]^{3/2}}, \  I_{10} = \frac{\tilde U^{IJK}\varepsilon_{IJK}}{[Y]^{3/2}}. 
\end{split}
\ee
In the above expressions a square bracket stands for trace on the flavor indexes and the spacetime indexes are contracted with the help of the full dynamical metric $g_{\mu\nu}$.~Our choice of basis is dictated by our interest in a slow roll phase with magnetic gaugid background: all the invariants, apart form $X$, are independent on the scale factor when evaluated on an FRW-gaugid background. $X$ can then be taken as the clock of cosmological evolution.
The most general gaugid lagrangian is thus a function of the above eleven invariants, 
\begin{equation} \label{gt}
{\cal L} = -Z\big(X, I_1, \dots, I_{10}\big)\, .
\end{equation}

\subsection{Friedmann-Robertson-Walker solutions}

In what follows, we'll be concerned with cosmological gaugid solutions, compatible with a spatially flat homogeneous and isotropic Friedmann Robertson Walker (FRW) background,
\begin{equation}
ds^2 = -dt^2 + a^2(t) d \vec x \cdot d \vec x\, .
\end{equation}
Furthermore, it is convenient to think in terms of the electric and magnetic fields $\vec E^I$ and $\vec B^I$. The only way a triad of spacelike vectors can be arranged to be compatible with homogeneity and isotropy is $E^I_j = f_1(t) \delta^I_j$ and $B^I_j = f_2(t) \delta^I_j$ for some two functions of  time $f_1$ and $f_2$. However, one of the Bianchi identities, $\dot {\vec B}^I + \vec \partial \wedge \vec E^I = 0$ immediately sets $f_2 = const$. In other words, up to rescaling, for a homoneneous solution one must have $B^I_j = \delta^I_j$. This means that the physical magnetic field $\vec B^I_{phys} = \vec B^I/a^2(t)$, for sufficiently homogeneous configurations, always scales as $a^{-2}$ in a FRW universe, independently of the lagrangian.

\vskip 0.15cm

Contrary to the case of the magnetic field, the scaling of $E^I$ depends on the chosen dynamics. To illustrate this, let us consider the simplest gaugid Lagrangian, ${\cal L} = -P(X)$. The equations of motion on FRW spacetime become $\partial_\mu(a^3 P'(X) F^{I\, \mu \nu}) = 0$. As expected, the constant magnetic component automatically drops out of the equations, and we are left with an equation constraining the dynamics of the electric field
\begin{equation}
\frac{d}{dt} (a P'(X) f_1) = 0\, .
\end{equation}
Although potentially interesting in their own right, \emph{electric gaugids}---the configurations with $f_1\neq 0$---will not be further explored in this paper. From now on we will focus on the \emph{magnetic gaugids}, defined by the background configuration~\eqref{op}. 

\vskip 0.15cm

One can argue that the magnetic gaugid is always a solution of the equations of motion as long as  the lagrangian respects parity $P$. Indeed, inspection of the building blocks in eq.~(\ref{realandim}) makes it clear that parity forces the lagrangian to depend only on even powers of the electric field.~Therefore, around a magnetic gaugid configuration with vanishing electric fields, the lagrangian is automatically stationary with respect to variations of the electric fields. Moreover on homogeneous backgrounds at hand, the action is obviosuly stationary with respect to variations of the $B_a^I$ as these are pure space derivatives. Hence the magnetic gaugid FRW universe is always a solution when $P$ is respected.
On the other hand, when $P$ is broken, there can exist a term linear in $E^I_a$ that  around a gaugid
cosmological background gives rise to a tadpole when varying with respect to the gauge field, according to the scheme
\begin{equation}
\delta \Big(a^\#(t) E B^n \Big) \ = \ - \ \partial_0 \Big(a^\#(t)  B^n\Big)\delta A_i 
\end{equation}
where we have integrated by parts. Because of the time dependence of the scale factor $a(t)$ a tadpole for the gauge field is generated in such a way that the solution 
for $A_\mu^I$ develops a time dependence and hence an electric field component.
For these reasons, we will be exclusively concerned with the magnetic gaugids~\eqref{op} in parity-preserving theories for the rest of the present study.

\vskip 0.15cm
 
The choice of the invariants~\eqref{invariants} has the advantage that, when considered on the background solution, only $X$ is sensitive to the scale factor: $X=24\cdot a(t)^{-4}$. In contrast, all other invariants $I_i$ are normalized in such a way to make them insensitive to the expansion of the universe.  
As a result, the stress tensor of the theory calculated on the background~\eqref{op} depends on these in a trivial way and takes on a simple form for the \textit{most general choice} of the function $Z$ in \eqref{gt} 
\be
\label{stresstensor1}
T^\mu_{~\nu} = -\delta^\mu_\nu Z + 4F^{~\mu}_{I~\alpha}F_{I\nu}^{~~\alpha}Z_X~,
\ee
where we have denoted $\partial Z/\partial X\equiv Z_X$.
The expressions for the energy density and pressure that follow from \eqref{stresstensor1} are
\be
\label{rhoandp}
\begin{split}
\rho &= Z \\
p &=-Z +\frac{4}{3} XZ_X~.
\end{split}
\ee
We will be particularly interested in a quasi-de Sitter background geometry which requires that the gaugid's energy density be approximately constant over an e-fold. Upon minimally coupling to gravity, the rate of variation of the Hubble parameter becomes
\beq
\label{eps}
\epsilon = -\frac{\dot H}{H^2} = \frac{2}{3}\frac{X Z_X}{\mpl^2 H^2}\ll 1~.
\eeq
Furthermore, the requirement that $\epsilon$ stay small ($\eta\equiv d\ln \epsilon / d\ln a \ll 1$) for at least $50-60$ e-folds as required by solving the horizon problem  translates into the following constraint 
\beq
\frac{d\ln Z_X}{d\ln X} = -1+\frac{\epsilon}{2}-\frac{\eta}{4}~,
\eeq
where we have used eq.~\eqref{eps} as well as the explicit expression for background VEV of $X$. This last equation yields a relation between the first and the second derivatives of $Z$,
\beq
\label{quasidscond}
\frac{X^2 Z_{XX}}{X Z_X} = -1 +\mathcal{O}(\epsilon, \eta)~,
\eeq
which we will use throughout to express $Z_{XX}$ in terms of $Z_X$. 

\vskip 0.15cm

In the rest of the paper we will explore a simple version of the theory, compatible with stability and (sub)luminality of the spectrum on cosmological (quasi-de Sitter) backgrounds. Extending our analysis to more general parity preserving lagrangians of the form \eqref{gt} is straightforward, but it is not expected to introduce any qualitative novelty compared to the minimal case we study in this paper. The theory we will focus on is specified by the following action
\beal
\label{themodel}
S = \int \dd^4 x \sqrt{-g}~\bigg[\frac{\mpl^2}{2}R -P(X)-(27 M^4_1 + 18M^4_2)~I_2 +72 M^4_2~ W\bigg] ,
\eeal
where we have defined
\beq
\label{yandw}
W \equiv \frac{F^{I~\beta}_{\alpha} F^{I~\gamma}_{\beta}F^{J~\delta}_{\gamma} F^{J~\alpha}_{\delta}}{X^2} = \frac{1+I_2+I_3}{4}~.
\eeq

\section{Kinematics, gauge redundancies and (an)isotropy}
\label{perts}

The system of three gauge fields under consideration clearly propagates six degrees of freedom on an arbitrary background. On  Minkowski  space  with vanishing  $F_{\mu\nu}^I$ background,  these modes form a triplet of massless helicity-1 representations of the Poincar\'e group, but on more general backgrounds they organize into irreducible representations of whatever internal/spacetime group remains unbroken. 
Furthermore, in the presence of dynamical gravity the multiplet structure is affected by the mixing with the metric perturbations. In this section, focussing on the magnetic gaugid phase \eqref{op}, we wish to explore the kinematics of perturbations, that is the structure of the multiplets and the convenient choices of gauge fixing. 
The residual space-time symmetry of this background, even in the presence of gravity on a FRW background, is the 3-dimensional euclidean group $ISO(3)$. At finite momentum $\vec k$, the perturbations are then classified according to their helicity, that is according to their transformation  under  $SO(2)$ rotations along $\vec k$.  At $\vec k =0$, on the other hand, and as we shall better elucidate, the physical perturbations must fall into representations of the full rotation group $SO(3)$.


\vskip 0.15cm


Before any gauge fixing is imposed, the gauge field perturbation\footnote{To simplify notation, we choose to put the flavor index as a subscript from now on.}, $a_{I\mu}\equiv A_{I\mu}-\langle {A}_{I\mu}\rangle$,
is given by a  general $3\times 4$ matrix. Under the residual $SO(3)$ symmetry of the background  \eqref{op}, flavor  and spatial indices are however identified. Therefore from now on we will not distinguish between the two, using lower-case letters also for the internal indices. The field $a_{I\mu}$ decomposes thus into a 3-vector $a_{i0}$ and a 3-tensor $a_{ij}$. It is important, however, to keep in mind that when considering the derivation of the electromagnetic fields, or the action of local diffeomorphisms, the first index ($i$) and the second ($0$ or $j$) play different roles. In particular, the magnetic field 
\beq
B_{im}=\epsilon_{mkj} \partial_k A_{ij}= \epsilon_{mkj} \partial_k (\epsilon_{ij\ell} x^\ell+a_{ij})\equiv - 2\delta_{im}+b_{im}
\eeq
is a two index tensor under the residual global $SO(3)$, but spatial diffs act only on the second index $m$.
%
Like  in ordinary gauge theories, the action does not involve time derivatives on $a_{i0}$. The $a_{i0}$ are therefore not dynamical,  their equations of motion are equivalent  to constraints on initial conditions and all propagating degrees of freedom live in $a_{ij}$.~At finite momenta, it is useful to decompose the latter quantity into helicity representations 
\be
\label{helicitydec}
a_{ij} = \delta_{ij}\alpha + \partial_i\partial_jS + \partial_iS_j + \partial_jS_i + E_{ij} + \epsilon_{ijk}\left(V_k + \frac{\partial_k T}{\sqrt{-\p^2}}\right) ~,
\ee
where $\p^2 \equiv \delta^{ij}\p_i\p_j$. Here, $\alpha$, $S$ and $T$ are helicity scalars, $S_i$ and $V_i$ are transverse helicity vectors ($\partial_i S_i=\partial_i V_i = 0$) and $E_{ij}$ is a symmetric, transverse and traceless helicity tensor mode ($E_{ii}=\partial_iE_{ij}=0$). In short, $\alpha$, $S$ and $T$ have helicity $h=0$,
$S_i$ and $ V_i $ have $h=\pm 1$, while $E_{ij}$ have $h=\pm 2$.
The $U(1)^3$ gauge redundancy  is given by $\delta a_{ij} = \partial_j \xi_i \equiv \partial_j(\xi^T_i+\partial_i\xi)$, where $\xi^T_i$ is the transverse part of the gauge parameter. One can then choose $\xi^T_i$ and $\xi$ so as to respectively  eliminate $S_i$ and $S$. In what follows we will mostly make this choice and parametrize  the spatial components of the gauge fields by
\be
\label{u1gaugefixed}
A_{ij} =\epsilon_{ijk}x^k+ \delta_{ij}\alpha + E_{ij} + \epsilon_{ijk}\left(V_k + \frac{\partial_k T}{\sqrt{-\p^2}}\right).
\ee

\vskip 0.15cm

Parity, defined by  $A_{I0}(t,\vec x)\to A_{I 0}(t,-\vec x)$ and  $A_{Ij}(t,\vec x)\to -A_{I j}(t,-\vec x)$, is preserved by the magnetic gaugid background \eqref{op} and can be used to classify the perturbations. As  $a_{ij}(t,\vec x)\to -a_{ij}(t,-\vec x)$, we then have that $T$ and $V_k$, in eq.~\eqref{u1gaugefixed}, are {\it polar} fields, i.e. with parity $=(-1)^h$ (in particular $T$ is a scalar and $V_k$ is a polar vector), while
 $\alpha$ and $E_{ij}$  are {\it axial},  i.e. with parity $=(-1)^{h+1}$. In particular, $\alpha$ and $E_{ij}$ are both parity odd.
The polar excitations $T$ and $V_k$ are  in fact  the exact analogs of the transverse and compressional phonons of an ordinary solid \cite{Endlich:2012pz}, but $\alpha$ and $E_{ij}$ are additional modes, specific to the gaugid. These extra phonons, in particular the helicity 2 mode $E_{ij}$, are the main source of novel cosmological effects.
%

\vskip 0.15cm

The above description of the gaugid modes is exact on flat space and in the absence of gravity. Moreover, even with dynamical gravity, it works approximately well at distances shorter than the cosmological horizon. This is just the usual short distance ``decoupling" of hydrodinamic and gravitational modes.
On the other hand, at distances comparable to the cosmological horizon and larger,
the effects of gravity must be included, in particular  its dynamical helicity-2  modes
must be incorporated. These, added to the gaugid modes, make up a total of eight physical helicity modes. It order to proceed
we will parametrize the metric in terms of the ADM variables,
\be
\label{adm}
ds^2 = -N^2 dt^2 + g_{ij}(dx^i+N^i dt)(dx^j+N^j dt)~,
\ee
where $N$ and $N^i$ are the \textit{lapse} and \textit{shift} variables, while $g_{ij}$ is the induced metric on uniform-time hypersurfaces. Diffeomorphism invariance can be used to fix various gauges. For the purposes of our analysis,  and depending on the precise question we'll be addressing, we shall employ two different gauge choices.

\vskip 0.15cm

For the purposes of studying the dynamics of finite momentum perturbations, we'll find it convenient to work in the \textit{spatially flat slicing gauge} (SFSG). In this gauge, the three-dimensional metric is only perturbed by the helicity-2 mode
\be
\label{spflat}
g_{ij}= a^2(t)\(e^{\gamma}\)_{ij}~,
\ee
where $\gamma_{ij}$ is the transverse traceless metric perturbation, satisfying 
\beq
\gamma_{ii} = \p_i\gamma_{ij}=0~.
\label{flatslicing}
\eeq
The rest of the modes that reside in the metric, in particular the lapse and the shift, are non-dynamical and can be algebraically integrated out from their respective equations of motion (the Hamiltonian and momentum constraints of general relativity). SFSG gauge  choice does not  affect the gaugid variables. This makes this gauge particularly convenient to study the  sub-horizon behaviour of the perturbations, where gravity and the gaugid decouple.

\vskip 0.15cm

An important question is that concerning the time evolution of superhorizon fluctuations, i.e.~those whose wavelengths exceed the Hubble horizon at or before the freezeout of the observable CMB modes. In particular, whether or not the isotropic magnetic gaugid is a dynamical attractor is closely related to this question. To address it, it is convenient to switch to the \textit{unitary} gauge, where the maximal possible number of propagating degrees of freedom is shifted from the gaugid sector into the metric. One can easily check that, by a proper choice of the \textit{spatial} coordinates, $V_i$ and $T$ in eq.~(\ref{u1gaugefixed}) can be eliminated. This way $a_{ij}$ is fully parametrized by the scalar $\alpha$ and the symmetric transverse-traceless $E_{ij}$ \footnote{Notice that $E_{ii}=\partial_jE_{ij}=0$ can  be imposed  at all $\vec k$'s, using the local $U(1)^3$.}
\beq
a_{ij}=\delta_{ij}\alpha+ E_{ij}~.
\label{unitary}
\eeq
There still remains to fix time diffs, i.e. choose constant time slices. The most convenient option\footnote{In section~\ref{reheating}, when discussing reheating, we shall find it convenient 
to choose $X$ as the clock.}, for the purpose of the present discussion, is to simply set $\gamma_{ii}=0$ in which case the spatial metric reduces to its trace free part $\gamma^T_{ij}$. 

\vskip 0.15cm

We are now ready to discuss the classical stability/instability of the magnetic gaugid cosmology with respect to initial small superhorizon inhomogeneities and anisotropies. Like in ordinary inflation, the kinematics around approximate de Sitter geometry ensures the dilution of any original small inhomogeneities. However the
possible anisotropies associated to deformations of the gaugid require a special discussion. As these perturbations will be streched to superhorizon scales during inflation, where they possibly become unobservable, the first basic question is: which homogeneous anisotropies are physically observable? In unitary gauge anisotropies can  arise from both the spacial metric $\gamma_{ij}^T $ and from the homogeneous magnetic fields associated to the residual gauge modes $\alpha$ and $E_{ij}$. In particular gauge field perturbations of the form
\beq
\alpha=\alpha_k x^k,\,\,\quad  E_{ij}=
E_{ijk}x^k
\label{constant}
\eeq
with $\alpha_k$ and $E_{ijk}$ constant tensors correspond to homogeneous anisotropic deformations of the background magnetic field. However not all these sources of anisotropies are physical, due to a residual gauge freedom that remains after imposing unitary gauge.  Consider indeed a homogeneous volume preserving change of spatial coordinates
\beq
x^i=D^i_{\,j} x'^j\simeq (\delta^i_{\,j}+\omega^i_{\,j})x'^j
\label{spacial}
\eeq
where $\omega^i_{\,j}$ describes the linearized transformation, satisfying $\omega^i_{\,i}=0$. The metric and magnetic field transform according to\footnote{The magnetic field tranforms like a contravariant vector (that is with $(D^{-1})^\ell_{\,j}$ instead of $D^\ell_{\,j}$). Indeed $B_{ij}=\epsilon_{jk\ell} 
B_{ik\ell}$ where $B_{ik\ell}\equiv \epsilon_{jk\ell} (\partial_ka_{i\ell}-\partial_\ell a_{ik}) $ is a covariant 2-form on the indices $k, \ell$. Moreover, given  
${\mathrm {det}}(D)=1$, one has $(D^{-1})^\ell_{\,j}=\epsilon^{\ell k m}\epsilon_{jpq}D^p_{\,k}D^q_{\,m}$, which explains the result in eq.~\eqref{shift}}
\beq
\begin{split}
g_{ij}(x,t)\to g_{k\ell}(D x,t)D^k_{\,i}D^\ell_{\,j} \quad&\Longrightarrow&\quad &\gamma_{ij}(x,t)\to\gamma_{ij}(x,t)+\omega_{ij}+\omega_{ji}\\
B_{ij}(x,t)\to B_{i\ell}(D x,t)(D^{-1})^\ell_{\,j}\quad&\Longrightarrow&\quad &b_{ij}(x,t)\to b_{ij}(x,t)+2 \omega_{ij}
\label{shift}
\end{split}
\eeq
which, as we shall better discuss at the end of this section, corresponds to fields still respecting the unitary gauge fixing ($\gamma^i_i=0$ is obviously preserved thanks to  $\omega^i_{~i}=0$, but for the gauge fields this is less obvious). 
Eq.~(\ref{shift}) immediately tells us that the most general homogeneous anisotropic magnetic field parametrized by a traceless matrix can be shifted into the spatial metric by a spacelike coordinate transformation: at long distances there is only one observable source of anisotropy described by the gauge invariant combination
\beq
\tilde \gamma_{ij}=\gamma_{ij} -\frac{1}{2}(b_{ij}+b_{ji}).
\label{obsanisotropy}
\eeq
Despite the existence of additional degrees of freedom,  the situation is then the same as in solid inflation, where in unitary gauge the only source of anisotropy is $\gamma_{ij}$. In practice, the physical anisotropy resides in the mismatch between the metric $\gamma_{ij}$ associated to the geometry of space and the effective  metric associated to the geometry of the medium, described respectively by
$\partial_i\phi^I\partial_j\phi^I$ for the solid and by $F^{I\mu}_{\,\,\,\,\,\,i}F^{I}_{\,\,\,\mu j}$ for the gaugid. This mismatch is most directly parameterized in terms of the diff invariant scalars $\partial^\mu\phi^I\partial_\mu\phi^J$ and $F^I_{\mu\nu}F^{J\mu\nu}$ for respectively solid and gaugid. One can check that the linear part of $F^I_{\mu\nu}F^{J\mu\nu}$,  with $I=i$ and $J=j$, is proportional to the combination in eq.~\eqref{obsanisotropy}.
Now, in solid inflation this physically observable mode decays $\propto a^{-\epsilon_\gamma}$ outside the horizon, where $\epsilon_\gamma$ is a slow roll parameter, essentially due to the presence of a `graviton mass'. 
%
As a result \cite{Bartolo:2013msa,Endlich:2013jia}, any initial anisotropy decays over a time scale of order $(\epsilon_\gamma H)^{-1}$. In particular, provided there are enough  \textit{total} inflationary e-folds ($N\gg \epsilon_\gamma^{-1}$), an order-unity classical and superhorizon $\gamma^T_{ij}$ at the beginning of inflation is driven to a negligible value by the time the observable CMB modes freeze out. As we shall show in the next section, also in gaugid inflation $\gamma_{ij}^T$ aquires a slow roll suppressed but positive mass.
The isotropic gaugid is therefore a (slow) attractor in full analogy with solid inflation.

\vskip 0.15cm

Before concluding this section we must however get back to the transformation property of the gauge field itself under the residual global diff. The result
obviously also has an impact on the form of the effective lagrangian for the perturbations. Now, the transformation in eq.~(\ref{spacial}) corresponds to
\beq
a_{ij}\to a_{ij}+(\epsilon_{ij\ell}\omega_{\ell k}+\epsilon_{i\ell k}\omega_{\ell j})x^k
\label{deltaaomega}
\eeq
which does not preserve the unitary gauge form of eq.~(\ref{unitary}). However, as we shall now show, one can perform a further gauge
transformation $a_{ij}\to a_{ij}+\partial_j\beta_i$ to bring the transformed field back to  unitary gauge. In order to proceed it is worth decomposing $\omega_{ij}$ into its  antisymmetric plus symmetric traceless parts, that is the spin 1 and spin 2 parts: $\omega_{ij}=\omega_{ij}^{(1)}+\omega_{ij}^{(2)}$, where $\omega_{ij}^{(1)}=\epsilon_{ijk}v_k$ and $\omega_{ij}^{(2)}-\omega_{ji}^{(2)}=\omega_{ii}^{(2)}=0$.
The spin 1 part, corresponds to an $SO(3)$ rotation of the space coordinates, for  which $\gamma_{ij}$ does not shift at linear order (see eq.~(\ref{shift})) while eq.~(\ref{deltaaomega}) gives
\beq
\delta^{(1)} a_{ij}=-\delta_{ij} v_k x^k+v_j x^i=-2\delta_{ij} v_k x^k+\partial_j(v_kx^kx^i)\, .
\eeq
We thus find that combining the $\omega^{(1)}$ with a gauge transformation with parameter $\beta_i=-v_kx^kx^i$ we obtain a tranformation $\tilde \delta^{(1)}$ that preserves the unitary gauge and defined by
\beq
\tilde\delta^{(1)}\gamma_{ij}=0,\qquad\tilde\delta^{(1)}\alpha= -2v_k x^k,\qquad \tilde\delta^{(1)} E_{ij}=0.
\eeq
The case of $\omega^{(2)}$ is slightly more involved, but one finds
\beq
\delta^{(2)}a_{ij}=\tilde \delta^{(2)}E_{ij}+\partial_j\beta^{(2)}_i
\eeq
with
\beq
\tilde \delta^{(2)}E_{ij}=\frac{2}{3}(\epsilon_{jk\ell}\omega^{(2)}_{ki}+\epsilon_{ik\ell}\omega^{(2)}_{kj}) x^\ell\,,\qquad\quad
\beta^{(2)}_i =\frac{1}{3}\epsilon_{i\ell m}\omega^{(2)}_{mk} x^\ell x^k
\eeq
where crucial use was made of the identity $\epsilon_{ij\ell}\omega^{(2)}_{\ell k}+\epsilon_{jk\ell}\omega^{(2)}_{\ell i}+\epsilon_{ki\ell}\omega^{(2)}_{\ell j}=0$. One thus finds that also for $\omega^{(2)}$ one can combine coordinate changes and gauge transformation to find a residual gauge tranformation that preserves the unitary gauge
\beq
\tilde\delta^{(2)}\gamma_{ij}=\omega^{(2)}_{ij}+\omega^{(2)}_{ji},\qquad\tilde\delta^{(2)}\alpha= 0,\qquad \tilde\delta^{(2)} E_{ij}=\frac{2}{3}(\epsilon_{jk\ell}\omega^{(2)}_{ki}+\epsilon_{ik\ell}\omega^{(2)}_{kj}) x^\ell .
\label{delta2}
\eeq
In the end, $\tilde \delta^{(1)}$ and $\tilde \delta^{(2)}$ correspond respectively to shifts in the spin 1 and spin 2 parts of the magnetic field $b_{ij}$. 
In addition to the residual $SO(3)$ and to the obvious constant shift symmetries of $\alpha$ and $E_{ij}$, the linearized lagrangian for the perturbations will then also be invariant under the tranformations $\tilde \delta^{(1)}$ and $\tilde \delta^{(2)}$.
Notice that the latter act on $\alpha$ and $E_{ij}$ as a generalization of galilean tranformations. In the case of $\tilde \delta^{(1)}$, given that only $\alpha$ transforms, 
no extra constraints than that placed by the constant shift of $\alpha$ arise: this field couples derivatively. $\tilde \delta^{(2)}$, on the other hand, dictates a specific relation  between the  mass term for $\gamma_{ij}$ and a term mixing $\gamma_{ij}$ with $\partial_kE_{ij}$, which is precisely what guarantees that the observable anisotropy slowly dilutes outside the horizon. 

In the above discussion, the use of UG had its logical convenience in the fact that  in this gauge the metric $\gamma_{ij}$ does not satisfy any differential constraints. 
On the other hand, the global residual tranformation in eq.~\eqref{delta2} also preseserves SFSG, since, given the constancy of $\omega^{(2)}_{ij}$, we have $\partial_i(\tilde\delta^{(2)}\gamma_{ij})=0$. Of course $\tilde\delta^{(2)}$ was built in such a way that $\tilde \delta^{(2)}V_i=
\tilde \delta^{(2)}T=0$ so that only $E_{ij}$ tranforms. In the computation of cosmological perturbations we shall work in SFSG and make use of this residual symmetry.

\vskip 0.15cm

One last comment concerns non linear orders. Clearly the residual coordinate change (\ref{spacial}) can be considered in its full non-linearity,  giving rise to a non-linear symmetry, that reduces at the lowest order to a galileon type transformation. The situation  will thus be similar to the case  of the conformal galileon, as concerns the tranformations of $\alpha$ and $E_{ij}$, though the tranformations of $\gamma_{ij}$ will have no analogue. In any case, this symmetry will definitely  provide constraints on the structure of the lagrangian at cubic and higher orders and play a role in the 
 study of non gaussianities.

\section{Dynamics of perturbations}
\label{section4}

In this section, we turn to exploring the \textit{dynamics} of the scalar, vector and tensor fluctuations in the gaugid phase \eqref{op}.  
We will start with introducing a representative model, free from all types of instability, as well as superluminal signal propagation for short-wavelength modes. We will then derive the expressions for the spectra of fluctuations of various helicity on quasi-de Sitter spacetimes, sourced by the magnetic gaugid. The main emphasis in this section will be on  the primordial \textit{tensor} spectrum in the model. The reason for that is twofold. First, as already remarked above, it turns out that the theory at hand offers a novel mechanism that can enhance the production of gravitational waves on inflationary quasi-de Sitter spacetimes. Second, we will find that  the rest of the theory---namely, the scalar and vector perturbations---lead to predictions that are identical to those of solid inflation. The cosmological perturbations in the latter model have been studied extensively by the authors of \cite{Endlich:2012pz}, and we will briefly review some of their results, as well as slightly extend their discussion of the scalar perturbations.

\vskip 0.15cm

It follows from eq.~\eqref{eps} that in order for the spacetime to deviate from perfect de Sitter, the action must necessarily depend on  $X$---the only invariant in our basis that depends on the scale factor. Thus, the simplest theory capable of describing a realistic inflationary scenario would correspond to a Lagrangian of the form $\mathcal{L}=-P(X)$, with $P$ some function of its argument (for the purposes of the present discussion, it is sufficient to focus on the flat-space/subhorizon theory).
A closer look at this minimal model reveals that the five modes associated to $\alpha$, $E_{ij}$ and $V_i$ are well behaved as long as the the quantity $XP_X$ is positive. (Notice that this is the statement of the \textit{null energy condition}: $XP_X$ is proportional to $\rho+p$ and upon coupling the system to gravity, positivity of this quantity implies an ever-decreasing Hubble rate $\dot H<0$, see eq. \eqref{eps}.) Furthermore, all of these five modes propagate at exactly luminal speeds. On the other hand, the sixth mode, the scalar $T$, turns out to be unstable: it has  a negative gradient energy, i.e. an imaginary speed of of propagation. This is a UV instability, and it is clear that taking into account mixing with gravity can not fix it.
However, a slight generalization of the minimal action does allow for fully stable and subluminal spectra, as we discuss next.


%

\subsection{A model with stable and subluminal perturbations}
\label{Amodel}

It is straightforward to show that the spectrum of a \textit{generic} low-energy theory of the magnetic gaugid possesses neither ghost,  nor gradient instabilities, nor superluminal modes. We provide a detailed proof of this statement in Appendix \ref{appa}. Rather than treating the problem in full generality, we present a simple model with a fully stable and subluminal spectrum. Extending our analysis to the most general case, while tedious in practice, is in principle straightforward.

\vskip 0.15cm

The theory we wish to explore in the remainder of this paper is specified by the action \eqref{themodel}.
In appendix \ref{appa1}, we study the spectrum of this theory in the short wavelength/subhorizon regime, where mixing with gravity becomes unimportant. We show that for the choice of the parameters $M^4_1$ and $M^4_2$ satisfying
\beal
\label{stcond}
XP_X>0~~ \bigwedge ~~ \frac{XP_X}{36}<M^4_1<\frac{XP_X}{18} ~~ \bigwedge ~~M^4_2>\frac{3M^4_1XP_X}{2XP_X-36M^4_1}~,
\eeal
all six degrees of freedom are fully stable and (sub)luminal. The short-wavalength $T,~V$ and $E$ modes propagate at the following velocities
\be
\label{speedsofsound}
c_T^2 = \frac{36 M_1^4 - X P_X}{3X P_X }~, \quad c_V^2 = \frac{(X P_X+18 M_1^4)(X P_X+12 M_2^4)}{X P_X (X P_X+24 M_2^4)}~,\quad c_E^2 = \frac{X P_X+36 M_1^4}{X P_X+24 M_2^4}~,
\ee
while the parity-odd scalar $\alpha$ has the following speed of sound: $c_\alpha^2 = c_E^2/(3c_T^2+2)$. Notice that consistency of the spectrum requires that neither of $M_1^4$ and $M_2^4$ vanish.

\vskip 0.15cm

It is straightforward to extend the calculation to include the effects of mixing with gravity.
From now on and until the end of this section, we will exclusively work in the spatially flat slicing gauge. To derive the quadratic lagrangian for the various fluctuations in the gaugid phase \eqref{op}, we go through the standard procedure of integrating out the non-dynamical degrees of freedom: the zeroeth components of the gauge fields $a_{i0}$, as well as the lapse and shift perturbations of the metric. This results in a non-local action for the polar modes, which is more convenient to write in terms of their Fourier components $T_{\vec k}=\(T_{-\vec k}\)^*$ and $V_{\vec k i} = (V_{-\vec k i} )^*$.\footnote{Our conventions for the Fourier transform are: $$T(x,t) = \int\frac{d^3k}{(2\pi)^3}~e^{i\vec k\cdot \vec x}~T_{\vec k}(t)~,$$ and similarly for $V_i$.} We leave the details of the derivation to appendix \eqref{appa}, and just quote the result for the full quadratic Lagrangian for the tensor, scalar and vector fluctuations:
\begin{align}
\label{actiontt}
S^{(2)}_{\rm TT} &=\int d^4x~ a^3~\bigg\{ \frac{\mpl^2}{8} \bigg( \dot \gamma^2_{ij}-\frac{1}{a^2}\(\p \gamma_{ij}\)^2 - 3 (c_T^2+1)H^2\epsilon \gamma_{ij}\gamma_{ij}\bigg ) \nn\\ 
&+\frac{(3 c_T^2+2)}{8}~\mpl^2 H^2\epsilon\(a^2c_E^{-2}\dot E^2_{ij}-(\p E_{ij})^2\)  + \frac{3}{4}(c_T^2+1)~\mpl^2 H^2\epsilon~\varepsilon_{ijk}\p_kE_{il}\gamma_{lj}\bigg\}~, \\
\label{Taction}
S^{(2)}_T &=\frac{\mpl^2}{4} \int dt \int \frac{d^3 \vec k}{(2\pi)^3}~ a^3~\bigg [ \frac{k^2/3}{1+k^2/3 a^2\epsilon H^2}\big|\dot T_{\vec k} + \epsilon H T_{\vec k}\big|^2-\epsilon H^2 c_T^2 k^2|T_{\vec k}|^2 \bigg]~, \\
\label{alphaaction}
S^{(2)}_\alpha &= \int ~ d^4 x ~a^3 ~\frac{3 c_T^2 + 2}{4 c_E^2}~\mpl^2 H^2\epsilon\bigg[ a^2\dot{\alpha}^2-c_\alpha^2  (\p\alpha)^2 \bigg] ~, \\
\label{vaction}
S^{(2)}_V &=\mpl^2 \int dt\int \frac{d^3 k}{(2\pi)^3} ~a^3~\bigg[ \frac{k^2/16}{1+k^2/16 a^2 \mathcal{N}_V^2}~ |\dot V_{\vec k i}|^2-c_V^2 \mathcal{N}_V^2 k^2 ~|V_{\vec k i}|^2  \bigg]~.
\end{align}
In the last formula for the vector action, we have defined
\beq
\mathcal{N}_V^2\equiv \frac{(2+3 c_T^2)}{4(2+3c_T^2+c_E^2)} H^2\epsilon~,
\eeq
which is a manifestly positive quantity. 

\vskip 0.15cm

The fields in the above Lagrangian describe $a_{ij}$ and $\gamma_{ij}$ in the SFSG, that is according to eqs.~\eqref{u1gaugefixed} and \eqref{flatslicing}.
We can however perform a suitable combination of diffs and gauge transformations, and use the same fields to parametrize $a_{ij}$ and $\gamma_{ij}$ in UG. The suitable tranformation is given by a diff $x^\mu \rightarrow {x'}^\mu = x^\mu + \xi^\mu$ combined with a gauge transformation $A_{ij}\rightarrow A_{ij} + \partial_j \beta_i$, where 
\begin{equation}\label{unitary-diff}
\xi^0 = \frac{\sqrt{-\partial^2}\, T}{6 H}, \qquad \xi^j = \frac12 \left(V^j+\frac{\partial^j T}{\sqrt{-\partial^2}}\right) \, ; \qquad \quad \beta_i=\epsilon_{ijk}\xi^j x^k\, .
\end{equation}
The end result is
\beq
\gamma_{ij}^{UG}=\gamma_{ij}^{SFSG}-\frac12\left(\partial_iV_j+\partial_j V_i\right)+\left (\frac{\partial_i\partial_j}{\partial^2}-\frac{\delta_{ij}}{3}\right)\sqrt{-\partial^2} T\, ,
\eeq
while $V_i$ and $T$ have simply disappeared from the gauge fields, 
\be
\label{aunitary}
A^{UG}_{ij} = \epsilon_{ijk} x^k + \delta_{ij}\alpha + E_{ij} .
\ee
In either SFGS or UG the above lagrangian is easily seen to satisfy the symmetry described by  eqs.~\eqref{shift} and \eqref{delta2}. The only transforming fields are $E_{ij}$ and $\gamma_{ij}$. All the terms, aside those involving no derivatives on  $\gamma_{ij}$, are obviously invariant under eq.~\eqref{delta2}.
In particular the variation of the $E_{ij}$ kinetic term is a total derivative, precisely like for the kinetic term of the Galileon. On the other hand the graviton mass and graviton-$E_{ij}$ mixing add up to
\beq
 - \frac{3\mpl^2}{8}\int d^4x~ a^3(c_T^2+1)H^2\epsilon\bigg( \gamma_{ij}\gamma_{ij}-2\varepsilon_{ijk}\p_kE_{il}\gamma_{lj}\bigg )
  \label{gammamass}
\eeq
where the expression in brackets reads $\gamma_{ij}\gamma_{ij}-2b_{ij}\gamma_{ij}$ in terms of the  magnetic field $b_{ij}$. Variation under eq.~\eqref{shift} is easily seen to be proportional to $b_{ij}$, which is a total space derivative. The above equation allows us to quickly deduce the time evolution of primordial homogeneous anisotropies in $\gamma_{ij}$ and $b_{ij}$. First of all, by a $\tilde\delta^{(2)}$ transformation any such anisotropy can be completely moved to the metric, in such a way that the constant mode of $b_{ij}$ vanishes. With such initial condition the constant mode of $\gamma_{ij}$ will evolve due to its mass term without inducing $b_{ij}$ back. That is because the $E-\gamma$ term induces a contribution $\propto \partial \gamma=0$ in the $E_{ij}$ equation of motion. As the graviton mass squared $3(c_T^2+1)H^2\epsilon$ is positive because of the requirements of microscopic stability of the gaugid,
the homogeneous mode will  decay at a rate $a^{-(c_T^2+1)\epsilon}$ and become negligible as soon as the total numer of e-folding $N_e> 1/\epsilon$. We here focussed on the evolution of the very constant mode, but, by continuity, the same reasoning applies to any primordial anisotropy mode of  wavelength comparable to Hubble, or longer, at the beginning of inflation.
\subsection*{A digression on notation}
\label{notation}

For readers' convenience, we collect in this subsection the definition of a number of parameters that will simplify the notation.
First off, we will find it convenient wo work in terms of the conformal time
\be
\tau = -\int_t^\infty \frac{dt}{a(t)}~.
\ee
Extra care will be required to account for the dependence of various quantities on the slow-roll parameters. In particular, one can make use of the \textit{exact} relation $d\(aH\)^{-1}/d\tau=\epsilon-1~,$
which, upon expanding $\epsilon$ around a reference conformal time $\tau_\star$ and integrating once, yields 
\beq
\label{taucorrected}
aH = -\frac{1+\epsilon_\star}{\tau}~,
\eeq 
where we have chosen the integration constant such that $a\to\infty$ when $\tau\to 0$. 

\vskip 0.15cm

Furthermore, it proves convenient to define the following (dimensionless) quantities
\beal
\label{invert}
\epsilon_{\gamma} =\frac{3}{2} (c^2_T+1)\epsilon ~, \qquad \epsilon_E=\frac{3 c_T^2+2}{c_E^2}\epsilon~,
\eeal
which are themselves weakly time-dependent:~$\eta_{\gamma,E} \equiv d \log \epsilon_{\gamma, E}/d\log a\ll 1$. With a bit of abuse of terminology, we will refer to all $\epsilon$- and $\eta$-parameters as `slow-roll' parameters. More explicitly, the $\epsilon$-parameters have the following approximate time dependence 
\be
\epsilon = \epsilon_{\star} \(\frac{\tau}{\tau_\star}\)^{-\eta_{\star}},\quad  \epsilon_{\gamma}= \epsilon_{\gamma\star}\(\frac{\tau}{\tau_\star}\)^{-\eta_{\gamma\star}}, \quad \epsilon_{E}= \epsilon_{E\star}\(\frac{\tau}{\tau_\star}\)^{-\eta_{E\star}}~.
\ee
We will set $\tau_\star$ to be the time corresponding to the longest CMB mode exiting the horizon during inflation, so that all modes of phenomenological interest cross the horizon at $|\tau|\leq |\tau_\star|$. The $`\star'$ subscript will indicate that the relevant quantity is evaluated at the time $\tau_\star$. Furthermore, we will assume that inflation ends $60$ e-folds after the longest CMB mode has left the horizon, 
\be
\frac{\tau_\star}{\tau_f}\equiv e^{N_e} \approx e^{60}~.
\ee

\vskip 0.15cm

For further reference, we quote the  expressions for the tensor and (parity even) scalar speeds of sound in terms of the $\epsilon$-parameters, obtained by inverting \eqref{invert}
\be
\label{tensorspeed}
c_E^2 = \frac{2\epsilon_\gamma-\epsilon}{\epsilon_E}~,\qquad c_T^2 = \frac{2\epsilon_\gamma-3\epsilon}{3\epsilon}~ .
\ee
These are also (weakly) time-dependent quantities: $c_E = c_{E\star}\(\tau/\tau_\star\)^{-s_{E\star}},~c_T = c_{T\star}\(\tau/\tau_\star\)^{-s_{T\star}}$~,
where $s_E$ and $s_T$ can be expressed in terms of the $\epsilon$ and $\eta$ parameters using eq.~\eqref{tensorspeed}:
\be
\label{sest}
s_E = \frac{1}{2}\bigg[\frac{2\epsilon_\gamma\eta_\gamma - \epsilon\eta}{2\epsilon_\gamma-\epsilon}-\eta_E\bigg]~,\qquad s_T = \frac{\epsilon_\gamma(\eta_\gamma-\eta)}{2\epsilon_\gamma-3\epsilon}~.
\ee

\vskip 0.15cm

In terms of the $\epsilon$-parameters, the stability and subluminality conditions \eqref{stcond} read
\beal
\label{stepsilon}
\epsilon>0~, \qquad \frac{3}{2}\epsilon<\epsilon_\gamma<2\epsilon~, \qquad \epsilon_E>\frac{\epsilon\epsilon_\gamma}{2\epsilon-\epsilon_\gamma}~.
\eeal
These yield, in particular, that the parameter $\epsilon_{\gamma}$ (which measures the graviton mass, as well as the $\gamma - E$ mixing in \eqref{actiontt}) is at most of order of the slow-roll parameter $\epsilon$, while $\epsilon_E$ can in principle be somewhat larger. Notice that $\epsilon_E\gg\epsilon$ corresponds to $c_E\ll 1$, as follows from eq.~\eqref{tensorspeed}.
The full set of independent slow-roll parametrs in the model under consideration thus consists of the three $\epsilon$'s, constrained by \eqref{stepsilon}, plus the corresponding $\eta$'s, which are free.

\vskip 0.15cm

Finally, for notational purposes, we introduce the following combination of  independent slow roll parameters:
\be
\label{bstar}
B \equiv -2\frac{\epsilon_\gamma}{\sqrt{\epsilon_E}}\equiv B_\star \(\frac{\tau}{\tau_\star}\)^{-\eta_\gamma + \eta_E/2}~,
\ee
with $B_\star \equiv -2\epsilon_{\gamma\star}(\epsilon_{E\star})^{-1/2}$. 

\vskip 0.15cm

Having made the above qualifications, we are ready for a closer discussion of the dynamics of the various helicities in gaugid inflation.

\subsection{Tensors}
\label{outofhorizon}

As already remarked above, the symmetry breaking pattern at hand does allow a mass term for the physical graviton, which is explicit in the quadratic action for tensor perturbations, eq.~\eqref{actiontt}.~Notice also the peculiar normalization of $E_{ij}$, whereby its kinetic term appears with an extra (strongly time-dependent) factor of $a^2$ compared to the graviton's kinetic term.\footnote{All other quantities in the lagrangian such as the various speeds of sound and the slow-roll parameters are only weakly time dependent on the quasi-dS spacetime under consideration.} This is a consequence of the approximate invariance under internal rescalings of the gauge fields $A^I_\mu \to \lambda A^I_\mu$, which in combination with diffeomorphisms leads to approximate symmetry under   $\{x^i\to \lambda x^i,~a\to \lambda^{-1} a,~E_{ij} \to \lambda E_{ij}\}$, left unbroken by the  background solution. The only explict souce of breaking of this approximate symmetry is the {\it weak} dependence of $P(X)$ on  $\langle X\rangle =24\cdot a^{-4}$, which implies a {\it weak} time dependence of all slow roll parameters.

\vskip 0.15cm

It is now  convenient to switch to conformal time and  to canonically normalized helicity-2 fields:
\beq
\label{cannorm}
\gamma_{ij}\to \frac{2}{a \mpl}\gamma^{c}_{ij}~, \qquad E_{ij}\to \frac{2}{a^2\mpl H}\frac{1}{\sqrt{\epsilon_E}} E^{c}_{ij}~.
\eeq
By eq. \eqref{taucorrected} and  by the definitions of the various $\epsilon$'s and $\eta$'s, the action for the tensor modes \eqref{actiontt} is then  written at leading order in the slow roll parameters as
\beal
\label{spin2canfin}
S^{(2)}_{\rm TT} &=\frac12 \int d^3x d\tau~ \bigg\{ \({\gamma^c_{ij}}'\)^2-\(\p \gamma^c_{ij}\)^2 + \frac{1}{\tau^2} (2+3\epsilon-2\epsilon_\gamma)\gamma^c_{ij}\gamma^c_{ij} \\ 
&+\({E^c_{ij}}'\)^2-c_E^2(\p E^c_{ij})^2 +\frac{1}{\tau^2} \(6+5\epsilon+\frac{5}{2}\eta_E\)  E^c_{ij} E^c_{ij}  \\  &-\frac{4}{\tau}\frac{\epsilon_\gamma}{\sqrt{\epsilon_E}}\epsilon_{ijk}\p_k E^c_{il} \gamma^c_{lj}\bigg\}~
\eeal
where a prime stands for derivation with respect to conformal time. It is convenient  to work in momentum space and expand in  helicity eigenstates \footnote{Given the helicity operator $s_{|| ij}\equiv i \hat k_l\varepsilon_{lij}$,  the polarizations eigenstates $\epsilon^s_{ij}(\vec{k})$ are defined by  $[s_{||},\epsilon^{\pm}] = \pm 2 \epsilon^{\pm}$. This property further implies transverse tracelessness, $k_i\epsilon^s_{ij}=\epsilon^s_{ii}=0$, and reflection hermiticity, $\epsilon^{s}(\vec k)^\star = \epsilon^{s}(-\vec k)$. The latter property, given the hermiticity of the spacetime fields, implies $\gamma^\dagger_s(\tau,\vec k ) = \gamma_s(\tau,-\vec k )$.}

\begin{align}
\begin{split}
\gamma^c_{ij} (\tau, \vec x) =  \int \frac{d^3 k}{(2\pi)^3}\sum_{s=\pm}~  \epsilon^s_{ij}(\vec{k})~ \gamma_s(\tau, \vec{k}) ~e^{i \vec{k}\cdot \vec{x}}\, , \\ 
E^c_{ij} (\tau, \vec x) = \int \frac{d^3 k}{(2\pi)^3}\sum_{s=\pm} ~ \epsilon^s_{ij}(\vec{k})~ E_s(\tau, \vec{k}) ~e^{i \vec{k}\cdot \vec{x}} \, .
\end{split}
\end{align}
In terms of the helicity eigenstates, $\gamma_\pm$ and $E_\pm$, the tensor action \eqref{spin2canfin} reads in  momentum-space\beal
\label{spin2canfin2}
S^{(2)}_{\rm TT} &=\frac12 \sum_{s=\pm} \int d\tau \frac{d^3 k}{(2 \pi)^3}~ \bigg\{ \gamma'_s (\vec k) \gamma'_s (\text{-}\vec k) - \left(k^2 - \frac{2+ 3\epsilon - 2 \epsilon_\gamma}{\tau^2}\right) \gamma_s (\vec k) \gamma_s (\text{-}\vec k) \\ 
&+ E'_s (\vec k) E'_s (\text{-}\vec k)  - \left(c_E^2 k^2 - \frac{6+ 5\epsilon + 5\eta_E/2}{\tau^2}\right) E_s (\vec k) E_s (\text{-}\vec k)   \\  &-\frac{2 B \, k }{\tau}  \left[\gamma_+({\vec k}) E_+(\text{-}{\vec k}) - \gamma_-({\vec k}) E_-(\text{-}{\vec k})\right] \bigg\}~.
\eeal
To avoid notational clutter, we have suppressed time dependence on the momentum modes and we have used the definition in \eqref{bstar}.

\vskip 0.15cm

The action~\eqref{spin2canfin2} describes four modes in total. However, fields of different helicity do not mix and can be treated separately. Their equations of motion are
\begin{align}
\label{gammeeq1}
\frac{d^2}{dz^2}\gamma_{\pm}+\(1-\frac{2+3\epsilon - 2 \epsilon_\gamma}{z^2}\) \gamma_{\pm}\pm  \frac{B}{z} E_{\pm}&=0~, \\
\label{gammeeq2}
\frac{d^2}{dz^2} E_{\pm}+\(c_E^2-\frac{6+5\epsilon + 5\eta_E/2}{z^2}\) E_{\pm}\pm  \frac{B}{z} \gamma_{\pm}&=0~,
\end{align}
where we have defined $z\equiv -k\tau$. Unbroken parity translates into invariance of the system \eqref{gammeeq1}-\eqref{gammeeq2} under $\gamma_{\pm}(\vec k)\to \gamma_{\mp}(-\vec k)$ and $ E_{\pm}(\vec k)\to- E_{\mp}(-\vec k)$.~

\vskip 0.15cm

In appendix \ref{gws} we give a detailed account of the quantization procedure for the system \eqref{spin2canfin2} on quasi-de Sitter space. The (momentum-space) fields of a given polarization are collected into two doublets $\phi_{\pm\alpha} = (\gamma_\pm, E_\pm)^T$, and each doublet is in turn expanded into two `eigenmodes' $\phi_{\pm\alpha} = \phi_{\pm\alpha}^{(1)}+ \phi_{\pm\alpha}^{(2)}$, where 
\be
\phi^{(n)}_{\pm\alpha} = f^{(n)}_{\pm\alpha}(\tau, \vec k) a_{\pm n} (\vec k) + f^{(n)}_{\pm\alpha}(\tau, \vec k)^* a^\dagger_{\pm n} (-\vec k)~.
\ee
Here $a_{\pm n}$ and $a_{\pm n}^\dagger$ are the corresponding annihilation/creation operators. 
The mode functions $f^{(1)}_{\pm}$ and $f^{(2)}_{\pm}$ solve the system in eqs.~(\ref{gammeeq1}, \ref{gammeeq2}). They can  be chosen such that $f^{(n)}_{-\alpha}=P_{\alpha\beta}f^{(n)}_{+\beta}$, where $P_{\alpha\beta}={\mathrm{diag}}(1,-1)$ represents parity.
Furthermore, provided they also obey certain orthonormality conditions spelled out in Appendix \eqref{gws}, 
the canonical commutation relations read
\be
[a_{\pm,m}(\vec{k}),a_{\pm,n}^\dagger(\vec{p})]=(2\pi)^3\delta^{(3)}(p-k)\delta_{mn}~, \qquad [a_{\pm,m}(\vec{k}),a_{\pm,n}(\vec{p})]=[a_{\pm,m}^\dagger(\vec{k}),a_{\pm,n}^\dagger(\vec{p})]=0\,.
\ee
For $c_E < 1$, one can choose modes with the following early time asymptotics
\begin{align}
\begin{split}
\label{ics}
f^{(1)}_{\pm}\(z\to \infty\)=\(
\frac{e^{iz}}{\sqrt{2k}},~ 0\)^T~,\quad 
f^{(2)}_{\pm}\(z\to \infty\)=\(0,~\pm\frac{e^{-i k \int d\tau c_E(\tau)}}{\sqrt{2c_E(\tau)k}}\)^T~\,.
\end{split}
\end{align}
corresponding to purely gravitational and purely gaugid Bunch-Davies excitations.

\vskip 0.15cm

At time $\tau$, the two-point function for the graviton's `$+$' polarization reads:
\be
\label{spectr}
\langle\gamma^c_+(\tau, \vec{k})~ \gamma^c_+(\tau, \vec{k'}) \rangle =(2\pi)^3\delta^3(\vec k+\vec k') \sum_{n} \big |f^{(n)}_{+1}(\tau,\vec k)\big|^2  \equiv (2\pi)^3\delta^3(\vec k+\vec k')P^+_\gamma~,
\ee
and similarly for the `$-$' polarization, with $P^-_\gamma=P^+_\gamma$ due to parity. The dimensionless tensor spectrum is thus given by 
\be
\label{tensorspectrum}
\Delta^2_t = \frac{k^3}{2\pi^2}\(\frac{2}{a\mpl}\)^2 \(P^+_\gamma + P^-_\gamma\) =\frac{k^3}{\pi^2} \(\frac{2}{a\mpl}\)^2 P^+_\gamma~,
\ee
where the second factor turns the canonically normalized fields into the physical ones. 

\vskip 0.15cm

Evaluating the $\langle\gamma \gamma \rangle$ two-point function at the end of inflation ($\tau = \tau_f$) thus boils down to evaluating the mode functions $f^{(1)}_+$ and $f^{(2)}_+$, with the initial conditions given in eq.~\eqref{ics} ($f^{(1)}_-$ and $f^{(2)}_-$ are then automatically obtained via a parity transformation). To put it another way, we have to solve the system \eqref{gammeeq1}-\eqref{gammeeq2} for $\gamma_+$, with the following two sets of initial conditions: $\{\gamma_+ \overset{z \rightarrow \infty}\longrightarrow  e^{iz}/\sqrt{2k},~ E_+ \overset{z \rightarrow \infty}\longrightarrow  0\}$ and $\{\gamma_+ \overset{z \rightarrow \infty}\longrightarrow  0,~ E_+ \overset{z \rightarrow \infty}\longrightarrow  e^{-i k \int d\tau c_E(\tau)}/\sqrt{2c_E(\tau)k}\}$.~We will see that the contribution of the first of the mode functions ($f^{(1)}_+$) to the amplitude of the $B$-mode power spectrum is at most of the order of the standard single field result \eqref{stdspec}, while the contribution of the second mode function ($f^{(2)}_+$) is parametrically larger. This results in a potentially significant enhancement of the primordial tensor spectrum. For these reasons, we will refer to the first and the second modes as the `small' and `large' modes respectively. 

\vskip 0.15cm

One can easily understand the origin of the enhancement of the gravitational waves by looking at the helicity-2 action \eqref{actiontt}, before canonically normalizing the fields. As we pointed out in section \ref{Amodel}  the residual symmetry $\tilde\delta^{(2)}$ forces the non-derivative part of the quadratic graviton lagrangian to have the form in eq.~\eqref{gammamass}, whose variation with respect to $\gamma_{ij}$ is proportional to
\be
\label{anisotropy}
\tilde\gamma_{ij}\equiv \gamma_{ij}- \frac{\epsilon_{ikl}\p_k E_{lj}+\epsilon_{jkl}\p_k E_{li}}{2}=\gamma_{ij}-\frac{b_{ij}+b_{ji}}{2}~,
\ee
which at large wavelengths precisely corresponds to the physical anisotropy perceived by local observers. 
Furthermore, and as we already argued in section \ref{Amodel}, given the graviton mass is positive,  $\gamma$ evolves in such a way that the anisotropy (that is the quantity in eq.~\eqref{anisotropy}) slowly decays in time. Schematically, 
\be
\label{gammatoE}
\gamma \overset{z \rightarrow 0}\longrightarrow \p E.
\ee
This is the situation as concerns classical fluctuations, but  $\p E$ also receives contributions from quantum fluctuations at {\it all} comoving wavelengths. During inflation these get constantly stretched to superhorizon scales, eventually freezing out and sourcing the gravitational waves according to eq. \eqref{gammatoE}. Modes satisfying eq. \eqref{gammatoE} are not observable as long as they are outside the horizon, but, like for all quantum fluctuations from inflation, we will eventually be concerned with the observable consequences after they re-enter the horizon. Now the crucial point here is that the gaugid's  physical excitations are characterized by a scale, smaller than $\mpl$, so that they undergo \textit{stronger} quantum fluctuations than gravitons do. In particular, inspection of the $E$ kinetic term in \eqref{actiontt} (see also eq.~\eqref{invert}) makes it clear that the quantum fluctuations in this mode are enhanced by a factor of $\epsilon_E^{-1/2}$ (in suitable units, to be specified below).~As implied by eq.~\eqref{gammatoE}, these large quantum $E$-fluctuations, once frozen out, source a  spectrum of super-horizon gravitational waves with amplitude
\be
\label{largemode}
\langle\gamma\gamma \rangle \sim \frac{H^2}{\mpl^2\epsilon_E}~.
\ee
The rest of the present section will be devoted to fixing the remaining factors in this formula.

\subsubsection*{The large mode}

We now show that the large mode  generates  a dominant contribution of the form~\eqref{largemode} to the amplitude of the primordial tensor spectrum. To this end, we need to solve the system \eqref{gammeeq1}-\eqref{gammeeq2} for $ \gamma_+$, with the following initial conditions
\be
\label{ic2'}
 \gamma_+\overset{z \rightarrow \infty}\longrightarrow 0~,\qquad  E_+\overset{z \rightarrow \infty}\longrightarrow \frac{e^{-i k \int d\tau c_E(\tau) }}{\sqrt{2c_E(\tau)k}}~.
\ee
The non-trivial profile of $\gamma_+$ is entirely due to the mixing with $ E_+$ (in the absence of this mixing $ \gamma_+$ would vanish at all times with the given initial conditions).

\vskip 0.15cm

For the rest of the argument, we will rely on perturbation theory in the small mixing parameter $B\sim \epsilon_\gamma/\epsilon_E^{1/2}$.
This parameter is at most of the order of $\epsilon_\gamma^{1/2}$, but can be smaller, consistently with stability of the theory, see eq.~\eqref{stepsilon}.
~We will argue, in particular, that $E_+$ evolves practically independently of $\gamma_+$ at all times, being well-approximated by the $B =  0$ solution of \eqref{gammeeq2}. To the linear order in all slow-roll parameters, this  solution reads 
\be
\label{Egr}
E_{+}=\(1+\frac{s_{E\star}}{2}\)e^{i (2\nu_E+1)\pi/4}\(\frac{\pi z}{4k}\)^{1/2} H^{(1)}_{\nu_E}\big(c_E(\tau)(1+s_{E\star})z\big), ~~ \nu_E = \frac{5}{2}+\epsilon_\star +\frac{1}{2}\eta_{E\star}+\frac{5}{2}s_{E\star},
\ee
where $H^{(1)}_{\nu_E}$ is the Hankel function of the first kind and we have appropriately normalized $E_+$ to match onto the early time asymptotics \eqref{ic2'}.
\vskip 0.15cm

One can straightforwardly verify that this expression indeed provides a good approximation to the exact solution of the system \eqref{gammeeq1}-\eqref{gammeeq2}. In particular, focussing first at early times ($z\gg 1$), the $E_+$ profile in \eqref{Egr} sources a non-trivial $\gamma_+$, which to the leading order in slow-roll takes the form
\be
\gamma_+ \big |_{z\gg 1}= - \frac{B_\star}{\sqrt{2c_{E\star}k} \, (1-c_{E\star}^2)}\frac{e^{i c_{E\star}z}}{z}~.
\ee
Plugging this expression back into eq.~\eqref{gammeeq2}, one can easily show that there is no significant backreaction on the early-time dynamics of gaugid's helicity-2 perturbations: the correction to the zeroth-order solution \eqref{Egr} scales as $\delta E_+/E_+|_{z\gg 1} = \mathcal{O}\(B_\star^2/z\)$.

\vskip 0.15cm

Next we explore the late-time dynamics of the system. The $z\ll1$ asymptotics of \eqref{Egr} read
\be
E_+\big|_{z\ll 1} = -\frac{3}{\sqrt{2k} c_{E\star}^{5/2}} ~\frac{1}{z_\star^{5 s_{E\star}/2}}~\frac{1}{z^{2+\epsilon_\star +\eta_{E\star}/2}},
\label{Eout}
\ee
where $z_\star \equiv -k\tau_\star$. Plugging this expression for $E_+$ into \eqref{gammeeq1} yields at late times
\be
\label{gammamastereq}
\frac{d^2  \gamma_{+}}{dz^2}-\frac{2+3\epsilon-2\epsilon_{\gamma}}{z^2}~ \gamma_{+} -  \frac{3B_\star}{\sqrt{2k}c_{E\star}^{5/2} }~\frac{1}{z_\star^{\eta_{E\star}/2+5s_{E\star}/2-\eta_{\gamma\star}}}~ \frac{1}{z^{3+\epsilon_\star+\eta_{\gamma\star}}}=0~,
\ee
where we have used eq. \eqref{bstar}.
The general solution to this equation is readily found:
\be
\label{gammalatetimes}
\gamma_+ =\frac{3 B_\star}{2\sqrt{2k}c_{E\star}^{5/2}\epsilon_{\gamma\star}}~\frac{1}{z_\star^{\eta_{E\star}/2+5s_{E\star}/2}}~ \frac{1}{z^{1+\epsilon_\star}}+C_+ z^{\lambda_+} +C_- z^{\lambda_-}~, \quad \lambda_{\pm}= \frac{1\pm (3+2\epsilon_\star -4\epsilon_{\gamma\star}/3)}{2}~,
\ee
where $C_+$ and $C_-$ are the two integration constants, which can in principle be determined by matching onto the early-time dynamics. 

\vskip 0.15cm

In solving the system \eqref{gammeeq1}-\eqref{gammeeq2}, we have not yet used the explicit expression for the mixing parameter $B$
in terms of the independent slow-roll parameters (recall that $B\propto\epsilon_\gamma/\epsilon_E^{1/2}$). Imagine for a second, therefore, that $B$ is an independent parameter and note the divergence of the first term in eq.~\eqref{gammalatetimes} in the limit $\epsilon_{\gamma}\to 0$, while keeping $B$ nonzero. However, at finite $z$, the original equations \eqref{gammeeq1}-\eqref{gammeeq2} and the approximation \eqref{gammamastereq} are  perfectly regular in this limit: this singularity should then be spurious and be removed, at finite values of $z$, once the boundary conditions are imposed. \footnote{The origin of the apparent singularity in eq.~\eqref{gammalatetimes} is  associated with the fact that the source term in \eqref{gammamastereq} has a scaling with $z$ that {\it{resonates}} with one of the solutions of the homogeneous equation. However we physically expect a resonance  should not give rise to singularities over short enough times, in line with our more mathematical argument that the solutions of \eqref{gammamastereq} should display no singularity in $\epsilon_\gamma$ at finite $z$.} This requirement alone automatically fixes the leading pieces in the expansion of $C_+$ and $C_-$ \textit{in the slow-roll parameters}:
\be
\label{cpcm}
C_+ = \mathcal{O}(1), \qquad C_- = -\frac{3 B_\star}{2\sqrt{2k}c_{E\star}^{5/2}\epsilon_{\gamma\star}}~\frac{1}{z_\star^{\eta_{E\star}/2+5s_{E\star}/2}} +\mathcal{O}(1)~.
\ee
The $C_+$ mode decays outside the horizon, so we will discard it from now on.  
Up to higher-order corrections in the $\epsilon$ and $\eta$ parameters, the late-time asymptotics of $\gamma_+$ for the large mode thus read
\be
\label{gammamastereq1}
\gamma_+ = -\frac{3}{\sqrt{2k}c_{E\star}^{5/2}\epsilon_{E_\star}^{1/2}}~\frac{1}{z_\star^{\eta_{E\star}/2+5s_{E\star}/2}}~\frac{1}{z^{1+\epsilon_\star}} \(1-z^{2\epsilon_{\gamma\star}/3}\),
\ee
where we have made use of the explicit expression for $B_\star$ in \eqref{bstar}. In this form it becomes evident that $\gamma_+$ vanishes in the limit $\epsilon_{\gamma}\to 0$, where the mixing between the two helicity-2 fluctuations disappears.

\vskip 0.15cm

One can already see that the expression \eqref{gammamastereq1} for the out-of-horizon profile of the graviton confirms our expectations, discussed around eqs.~\eqref{anisotropy}-\eqref{largemode}, concerning the sourcing of gravitational waves by the gaugid's helicity-2 mode. First of, notice the enhancement of the wavefunction by a factor of $\epsilon_E^{-1/2}$, in agreement with \eqref{largemode}. Furthermore, the first term in \eqref{gammamastereq1} describes a constant profile for the \textit{physical} graviton (see eq.~\eqref{cannorm}), sourced by the frozen out-of-horizon $E_+$ modes. The second term, on the other hand, describes the slow dilution of physical anisotropy due to the graviton mass, $m_\gamma^2 = 2 H^2\epsilon_\gamma$, as discussed before eq.~\eqref{gammatoE}.

\vskip 0.15cm

The profile \eqref{gammamastereq1} for $\gamma_+$  leads to negligible backreaction once plugged back into the $E_+$-equation of motion \eqref{gammeeq2}. Indeed, the source term in the resulting equation, being of order $\mathcal{O}\(B^2z^{-2}\)$, is $O(B^2)$ with respect to the unperturbed solution at horizon crossing $z=O(1)$, while it is even more
 negligible at late times compared to leading terms that grow like $\mathcal{O}\( z^{-4}\)$. This means in particular, that there is no resonant effect similar to the one that arises in $E_+$-sourcing of the gravitational waves. The leading effect therefore originates from effects at horizon crossing 
 and corresponds to an $O(B^2)$ correction to the asymptotic behaviour of $E$ given in  eq.~\eqref{Eout}.  In turn this will similarly, and negligibly, modify \eqref{cpcm}
by effects of relative size $O(B^2)$.

\vskip 0.15cm

From the above discussion it is clear that the $\gamma_+$ backreaction on the dynamics of the gaugid's helicity-2 mode is small ($O(B^2)$) at all values of $z$. We have explicitly verified this fact via a numerical integration of the system \eqref{gammeeq1}-\eqref{gammeeq2}. 

\vskip 0.15cm

The perturbative expansion in the small mixing parameter $B$ is thus at work as far as the large mode is concerned. Next we explore the status of perturbation theory for the small mode. 

\subsubsection*{The small mode}

Evaluating the contribution of $f^{(1)}_+$ to the primordial B modes amounts to solving the system \eqref{gammeeq1}-\eqref{gammeeq2} for $\gamma_+$, with the following initial conditions:
\be
\label{ics3}
\gamma_+ \overset{z \rightarrow \infty}\longrightarrow \frac{e^{iz}}{\sqrt{2k}}~,\qquad  E_+\overset{z \rightarrow \infty}\longrightarrow 0~.
\ee
In line with the anticipated perturbative expansion in the small mixing parameter $B$, we expect $E_+ = \mathcal{O}(B)$ at all times, while $\gamma_+$ is given by the following expression
\ba
\label{gammagr'}
\gamma_{+}=e^{i (2\nu+1)\pi/4}\(\frac{\pi z}{4k}\)^{1/2} H^{(1)}_\nu(z) + \delta\gamma_{+}~,
\ea
where $\delta\gamma_+ = \mathcal{O}(B^2)$. 
Here $\nu \equiv 3/2+\epsilon_1/3$, and the solution has been appropriately normalized to match onto the Bunch-Davies initial condition for $\gamma_+$.

\vskip 0.15cm

To compute the correction term in \eqref{gammagr'}, consider first the early-time ($z\gg 1$) dynamics of the system. With the initial conditions set by eq. \eqref{ics3}, the non-trivial $E_+$ profile is entirely due to the mixing with the graviton, and for $z\gg 1$ it becomes
\be
\label{eearly}
 E_+ \big |_{z\gg 1}\approx ~\frac{B}{\sqrt{2k} (1-c_E^2)}\frac{e^{i z}}{z}~.
\ee
Plugging this expression back into the equation for $\gamma_+$, one finds  
\be
\delta \gamma_+\approx -i \frac{B^2}{2\sqrt{2k}(1-c_E^2)}~\frac{e^{iz}}{z}~.
\ee
The corretion to $\gamma_+$ due to mixing with the gaugid's helicity-2 excitation is indeed small at early times, $\delta\gamma_+ = \mathcal{O}(B^2/z)$.

\vskip 0.15cm

The late-time dynamics is more subtle, however.~Indeed, according to \eqref{gammagr'}, the zeroth-order graviton wavefunction grows like 
\be
\label{gamma''}
\gamma_+\approx \frac{i}{\sqrt{2k}} \frac{1}{z^{1+\epsilon_1/3}}~
\ee
in this regime. When plugged into the $E_+$ equation of motion \eqref{gammeeq2}, this acts as a source, giving rise to the following late-time profile
\be
\label{e''}
E_+\big|_{z\ll 1}\approx  i \frac{B}{6\sqrt{2 k}}\frac{1}{z^{\epsilon_\star - 2\epsilon_{\gamma\star}/3}} + D_+ z^{\kappa_+} + D_-z^{\kappa_-}, \quad \kappa_{\pm} = \frac{1\pm (5+2\epsilon_\star+\eta_{E\star})}{2}~.
\ee
Here $D_{\pm}$ are the two integration constants. One generically expects the corresponding modes to be both excited at horizon crossing $z\sim 1$, with amplitudes $D_{\pm}\sim B$.

\vskip 0.15cm

Perturbation theory is only at work if \eqref{e''} does not lead to a significant backreaction on $\gamma_+$ when plugged back into \eqref{gammeeq1} as a source. For $z\ll 1$, it is clear that the first two terms in \eqref{e''} have no effect on the late-time dynamics of the graviton. The last term, however, scales as $z^{-2-\epsilon_\star - \eta_{E\star}}$, and we have seen that this is precisely the kind of scaling that results in a resonant enhancement of the gravitational waves ouside the horizon. Going through the same analysis as for the large mode, we thus have
\be
\frac{\delta \gamma_+}{\gamma_+}\big|_{z\ll 1} \sim \frac{B D_-}{\epsilon_{\gamma}} \sim \frac{B^2}{\epsilon_{\gamma}} = 4\frac{\epsilon_{\gamma}}{\epsilon_{E}}.
\ee
This can be considered a small perturbation only for $\epsilon_{E}\gg \epsilon_\gamma$ (recall that this part of the parameter space is allowed by stability constraints \eqref{stepsilon}). Otherwise, the effects from the backreaction of $E_+$ on the late-time dynamics of $\gamma_+$ are $O(1)$. 

\vskip 0.15cm

However, even when $\epsilon_E\sim \epsilon_\gamma$, and the contribution of the backreaction on the small mode cannot be computed by treating the mixing $B$ as a perturbation, it remains true that  the contribution of this mode to the primordial spectrum of gravity waves remains of the standard magnitude $\Delta_t^2\sim H^2/\mpl^2$. This observable is therefore fully dominated by the large mode, which we have been able to compute reliably in perturbation theory. For that reason, we will discard the sub-dominant small mode from now on. 

\vskip 0.15cm
Using \eqref{gammamastereq1} as well as the relation $a(\tau)=a(\tau_\star)\(\tau_\star/\tau\)^{1+\epsilon_\star}$ in eq.~\eqref{tensorspectrum}, one can straightforwardly evaluate the expression for the tensor power spectrum at the end of inflation $\tau = \tau_f$.  We will reproduce the result for the two limiting cases of a not-so-small $\epsilon_\gamma$ as well as a very small one, $\epsilon_\gamma \ll N^{-1}_e \approx 1/60$.~The primordial B mode spectrum takes the following form in these limits: 
\begin{equation}
\label{ttspectrumtm}
  \Delta^2_t (\tau_f) =  \left \{
  \begin{aligned}
    &\frac{18}{\pi^2}~ \frac{H^2_\star}{\mpl^2}~\frac{1}{c_{E\star}^5 \epsilon_{E\star}}~ \frac{1}{(-k\tau_\star)^{2\epsilon_\star+\eta_{E\star}+5s_{E\star}}},~~ \epsilon_{\gamma} N_e\gsim 1  \\
 &\frac{8}{\pi^2}~ \frac{H^2_\star}{\mpl^2}~\frac{\epsilon^2_{\gamma\star}}{c_{E\star}^5 \epsilon_{E\star}}~ \frac{1}{(-k\tau_\star)^{2\epsilon_\star+\eta_{E\star}+5s_{E\star}}}\log^2(-k\tau_f),~~  \epsilon_{\gamma} N_e\ll 1 
  \end{aligned} \right. 
\end{equation}  
Assuming for concreteness that $\epsilon_{\gamma} N_e\sim 1$, we have:
\be
\Delta^2_t\sim \frac{H_\star^2}{\mpl^2}~\frac{1}{c_{E\star}^5 \epsilon_{E\star}}.
\ee
This exceeds (especially for a somewhat suppressed $c_{E\star}$) the standard single-field result \eqref{stdspec} by a large factor $c_{E\star}^{-5}\epsilon_{E\star}^{-1}$. 

\vskip 0.15cm

One may wonder to what extent is such an enhancement of the tensor modes in gaugid inflation compatible with the current limits on the tensor-to-scalar ratio. We will see below that the spectrum of scalar perturbations is given by a similar expression in the model at hand
\be
\label{giscspec}
\Delta^2_s \sim \frac{H_\star^2}{\mpl^2}\frac{1}{c_{T\star}^5 \epsilon_\star}~.
\ee
For a somewhat small $c_T$ compared to $c_E$, the tensor-to-scalar ratio is thus easily suppressed beyond the observational upper limit. Another possibility for suppressing gravity waves relative to scalar CMB fluctuations is to have $\epsilon_\gamma \ll N_e^{-1}$.

\vskip 0.15cm

The \textit{tilt} of the tensor spectrum can be readily read off eq. \eqref{ttspectrumtm}:
\begin{equation}
\label{ttilt}
 n_t =  \left \{
  \begin{aligned}
    & - 2\epsilon_\star - \eta_{E\star} - 5 s_{E\star},~~~\epsilon_{\gamma} N_e\gsim 1 \\
    &-\frac{2}{N_e} - 2\epsilon_\star - \eta_{E\star} - 5 s_{E\star},~~~\epsilon_{\gamma} N_e\ll 1
  \end{aligned} \right. 
\end{equation} 
where $s_E$ has been defined in \eqref{sest}.
In both of the above limits one expects a percent-level tensor tilt, $n_t\sim N_e^{-1}$, which appears to be a genuine prediction of the theory. For all slow-roll parameters smaller than $ N_e^{-1}$, the tensor spectrum is red-tilted, while in a more general case both signs of $n_t$ are possible. 

\vskip 0.15cm

One last comment concerns the amount by which the $\gamma_+$ spectrum \eqref{ttspectrumtm} evaluated right before the end of inflation varies by the time the CMB modes reenter the horizon. We will return to this question in section \ref{reheating}. Generically, the predictions for the primordial tensor and scalar spectra are expected to be more sensitive to the details of reheating in the model at hand, than they are in more ordinary theories of inflation. We will nevertheless argue, that at least in the case that reheating happens fast, i.e. within a single Hubble time, the asymptotic tensor spectrum is reproduced by \eqref{ttspectrumtm} to a good approximation.

\subsection{Scalars}\label{scalarsec}

The scalar sector of gaugid inflation consists of a pair of dynamical degrees of freedom.~Importantly, these have opposite parity with respect to the unbroken symmetry under inversion of spatial coordinates: $\alpha$ is parity-odd, while $T$ is parity even. In the parity-symmetric theory at hand, this results in a complete decoupling of  the former field both from the metric and from the parity-even mode at the quadratic order in the perturbation lagrangian, making the primordial scalar spectrum sensitive to $T$ alone. 

\vskip 0.15cm 

To compute the action for $T$ we must first solve the contraints associated to the non-dynamical fields. By parity, the only fields that can
affect the action for $T$ must also be scalars.
In the spatially flat slicing gauge, the only such fields reside in
in the lapse and the shift variables
\be
N = 1+\delta N~, \qquad N_i=\p_i\psi~~~ (N^i = g^{ij} N_j)~,
\ee
while the scalar residing in the gauge field $a_{i0}=\partial_i \chi$ is indeed a pseudo-scalar, and so it does not mix (see App.~\ref{appa}).
The complete perturbation lagrangian for the scalar modes is somewhat complicated, and is reproduced in Appendix \ref{scalarmixing}.
Passing to momentum space, the lapse and shift can be integrated out from their respective equations of motion (the Hamiltonian and mometum constraints of general relativity), which yields
\be
\label{lapseandshift}
\delta N = - \frac{a^2 \dot H}{2 k H}~ \frac{\dot T - \dot H T/H}{1 - 3 a^2\dot H/k^2}~, \qquad \psi = -\frac{a^2}{2k} ~ \frac{3 a^2 \dot H \dot T/k^2 -\dot H T/H}{1 - 3 a^2 \dot H/k^2}~.
\ee
Plugging these expressions back into the action leads to the final form of the quadratic $T$ Lagrangian in eq.~\eqref{Taction}.

\vskip 0.15cm

As advertised, this is the exact same action describing the only scalar mode in solid inflation. It is not difficult to see why this is the case by taking a brief detour to the unitary gauge, where $T$ is `eaten' by the metric and the gauge fields contain only the parity odd scalar and the tensor modes, see eq.~\eqref{aunitary}. If writing down the quadratic action for the parity even scalar is our only concern, we can calculate the  invariants in~\eqref{invariants} directly with the \emph{unperturbed} gauge fields, 
$A_{ij} = \epsilon_{ijk}x^k$ but with a perturbed metric. (We have already explained that the auxilliary fields in the gauge sector do not matter because of their quantum numbers.) By inspection, we notice that in unitary gauge, our invariants reduce to various contractions of the spatial contravariant metric $g^{ij}$, with the Kronecker $\delta_{ij}$ contracting the indices. For example, we have in obvious notation
\begin{equation}
X^{UG} = 4\left([g]^2 - [g^2]\right) + \dots \, , \qquad I_2^{UG} = 2 \frac{[g^2]^2 - [g^4]}{([g]^2 - [g^2])^2} + \dots \, ,
\end{equation}
where ellipses stand for pieces containing helicity/parity modes, other than the parity even scalar. Since $g^{ij}$ is formally a 3 by 3 matrix, invariants higher than $[g^3]$ can be expressed as functions of $[g]$, $[g^2]$ and $[g^3]$. But these are precisely the ingredients of the solid lagrangian written in the unitary gauge~\cite{Endlich:2012pz}, where the spatial coordinates are chosen to lie along the comoving coordinates of the solid, $x^i = \delta^i_J \phi^J$. We will see shortly that it is precisely $T$---our only parity-even scalar mode---whose dynamics determine the properties of the physical curvature perturbations in gaugid inflation. Given the full analogy to the scalar sector of solid inflation studied in ref.~\cite{Endlich:2012pz}, we will briefly summarize various points addressed by the authors of that reference as well as slightly extend their discussion. 

\vskip 0.15cm

The (scalar part of the) gaugid's perturbed stress tensor  is calculated in appendix \ref{strten}. Away from any gauge, we define the perturbed metric as follows
\be
\label{metricdef}
ds^2 = -(1+2 A)dt^2 + 2a\p_i B dx^i dt + a^2 \big [(1-2 C)\delta_{ij}+2\p_i\p_j D \big ]dx^idx^j~, 
\ee
while our definitions for the comoving curvature perturbation $\mathcal{R}$ and the curvature perturbation on uniform density surfaces $\zeta$ are: 
\be
\label{zetardef}
\begin{split}
\mathcal{R} &= C - \frac{H}{\rho + p}\delta q, \\
\zeta &= C  - \frac{\delta\rho}{3(\rho+p)},
\end{split}
\ee
where $\delta q$ is the potential for momentum density.

\vskip 0.15cm

In SFSG, the two gauge-invariant curvature perturbations read
\beal
\label{randz}
\mathcal{R}_{\vec k} &= \frac{k}{6 H\epsilon}~\frac{\dot T_{\vec k}+H\epsilon T_{\vec k}}{1+k^2/3a^2 H^2\epsilon}~, \\
\zeta_{\vec k} &= \frac{k}{6} T_{\vec k}~.
\eeal
One peculiarity of the theories under consideration is that $\mathcal{R}$ and $\zeta$ are neither equal nor conserved on superhorizon scales during the inflationary phase: they slowly evolve according to
\be
\label{zetadot}
\dot\zeta \sim \epsilon H\zeta~,
\ee 
and similarly for $\mathcal{R}$.
(See, e.g., the expression \eqref{scalarspec} for the $\langle\zeta\zeta \rangle$ two-point function below.) This is in stark contrast to more conventional theories of inflation, where Weinberg's theorem \cite{Weinberg:2003sw} guarantees the presence of a pair of adiabatic modes, one of which has constant and identical $\mathcal{R}$ and $\zeta$ at superhorizon scales, while for the other one both of these curvature perturbations vanish. The origin of the non-compliance of scalar perturbations of solids/gaugids to Weinberg's theorem can be traced back to these systems' anisotropic stress $\sigma_{ij}$ \cite{Endlich:2012pz}, which remains sizeable even in the long-wavelength limit: one of the key technical assumptions of the theorem is that $\sigma_{ij}$ decays at $k/a\to 0$\cite{Weinberg:2003sw}.

\vskip 0.15cm

Another assumption of Weinberg's theorem is that it is possible to continue all homogeneous perturbations to physical ones at finite momentum.~However, this assumption does not hold in solid/gaugid inflation either. 
Indeed, let us try to see the origin of the uncommon features of $\mathcal{R}$ and $\zeta$ directly from Einstein's equations.~Equality of $\mathcal{R}$ and $\zeta$ as well as their conservation outside the horizon are usually established based on the two gauge-invariant equations (see, e.g., \cite{Bassett:2005xm})
\ba
\label{zetar}
\zeta - \mathcal{R}  &=&  \frac{1}{3 H^2\epsilon}\frac{k^2}{a^2}\Psi \\ 
\label{zetadot0}
\dot\zeta &=& -\frac{1}{3 H}\frac{k^2}{a^2} \Psi + \dots~,
\ea
where the Bardeen potenial $\Psi$ is given by the following expression in the spatially flat gauge\footnote{In an arbitrary gauge, it is defined as $\Psi = C + a^2 H (\dot D -B/a)$~.}
\be
\label{bardeen}
\Psi = - H\psi~,
\ee
and the ellipses denote extra terms that do not play a role in the argument to come.~Equality and constancy at large scales of the two curvature perturbations requires that the quantity $(k^2/a^2)\Psi$ vanish at zero momentum. This is usually true in conventional inflationary theories, where the Bardeen potential is regular for $k/a\to 0$.~In contrast, in solid/gaugid inflation, $\Psi$ diverges as $(k/a)^{-2}$ in this limit, so that $(k^2/a^2)\Psi$ is finite.~To see this, note that the explicit expressions \eqref{lapseandshift} for the curvature perturbations imply the following (schematic) relations at small momenta: $\zeta \sim \mathcal{R} \sim k T $. In the same limit, eqs.~\eqref{lapseandshift} and \eqref{bardeen} yield 
\be
\frac{k^2}{a^2}\Psi \sim Hk\dot T\sim \epsilon H^2\zeta~, 
\ee
where to establish the last relation we have used \eqref{zetadot}.
The effects from the $\Psi$-dependent terms are therefore only slow-roll suppressed, but do not decouple as $k/a\to0$ in either of the (gauge-invariant) equations \eqref{zetar} and \eqref{zetadot0}, usually used to prove that $\mathcal{R}$ and $\zeta$ are identical and conserved outside the horizon.

\vskip 0.15cm

Just as for the tensor modes, non-coincidence and slow time evolution of $\zeta$ and 
$\mathcal{R}$ outside the horizon makes the predictions of the model at hand somewhat sensitive to the details of reheating.~In particular, there are two immediate questions: i) given the non-zero time derivative of the two superhorizon curvature perturbations at the and of inflation, how much do their values at $\tau = \tau_f$ differ from their measured value at CMB decoupling?  ii) out of the two perturbations $\zeta(\tau_f)$ and $\mathcal{R}(\tau_f)$ that differ outside the horizon \textit{immediately before the end} of inflation, which combination, if any, have we measured in the CMB? 
As a matter of fact, we will see in the next section that $\mathcal{R}$ is generically expected to change discontinuously across the reheating surface, while $\zeta$ is continuous \cite{Endlich:2012pz}. For that reason, we will proceed assuming that it is the latter curvature perturbation that determines the observed features of the CMB. Moreover, we will argue in the next section, that the non-zero variation rate of $\zeta$ at the end of inflation cannot make its asymptotic, measured value too different from $\zeta(\tau_f)$, as far as reheating happens fast.

\vskip 0.15cm

Making the identification $\zeta=kT/6$ along the lines of the previous discussion, the two-point function of $\zeta$ has been calculated based on the action \eqref{Taction} in ref.~\cite{Endlich:2012pz}, and the result reads
\beal
\label{scalarspec}
\langle \zeta(\tau,\vec k_1) \zeta(\tau,\vec k_2)\rangle \bigg |_{k\tau\to 0^-}=(2\pi)^3 \delta^3(\vec k_1+\vec k_2)~\frac{H_\star^2}{4\epsilon_\star c_{T\star}^5\mpl^2}~\frac{1}{k_1^3}~\frac{(\tau/\tau_*)^{8 c^2_{T\star}\epsilon_\star/3}}{(- k_1\tau_*)^{5 s_{T\star}-2 c_{T\star}^2\epsilon+\eta_\star}}~.
\eeal
The tilt of the spectrum can be readily read off of this expression:
\be
\label{scspec}
n_S - 1 \ \simeq \ 2 c_{T\star}^2\epsilon_\star -5s_{T\star}-\eta_\star \, ,
\ee
where the parameter $s_T$ has been defined in \eqref{sest}.
Phenomenologically, the scalar tilt can be fixed to the observed value by e.g. dialling the free parameter $\eta_\star$, which does not enter into the expression \eqref{ttilt} for the tensor spectrum.

\vskip 0.15cm

Recalling the form of the tensor spectrum \eqref{giscspec} and using eq.~\eqref{invert} one finds that the tensor-to-scalar ratio is given for $\epsilon_\gamma\lsim N^{-1}_e$ by the following expression
\begin{equation}
\label{r}
 r \sim \frac{c_T^5}{c_E^3}\times  (\epsilon_\gamma N_e)^2~.
\end{equation}
For $\epsilon_\gamma$ somewhat larger than $N^{-1}_e$, on the other hand, the expression for $r$ is saturated at $r\sim c_T^5/c_E^3$.
One can see that $r$ is extremely sensitive to $c_T$ and can thus be highly suppressed if the scalar phonons are subluminal, $c_T<1$. As can be straightforwardly inferred from \eqref{stepsilon}, the stability and (sub)luminality constraints do allow for this possibility. In particular, dialling the value of $\epsilon_{\gamma}$ to lie slightly above $3\epsilon/2$ (thus saturating the lower bound on $\epsilon_\gamma$ in \eqref{stepsilon}), the scalar speed of sound, $c_T$, tends to zero, while the tensors propagate at a finite speed, approaching $c_E^2\simeq 2/3$ as the last condition in \eqref{stepsilon} becomes saturated, see eq. \eqref{tensorspeed}. Close to this point in the parameter space, $r$ can be arbitrarily small.~On the reverse side, the stability/subluminality constraints do allow for a \textit{large} tensor-to-scalar ratio as well. In fact, $c_E$ can be arbitrarily suppressed compared to $c_T$ (and $r$ can thus be made arbitrarily large) by choosing, for fixed $\epsilon$ and $\epsilon_\gamma$, a large enough parameter $\epsilon_E$ in \eqref{tensorspeed}.\footnote{Note that $c_T$ has no dependence on the latter parameter at all.} Again, it is clear from \eqref{stepsilon} that such a choice of the parameters of the theory does not clash with stability and/or (sub)luminality of the spectrum.

\vskip 0.15cm

Notice, finally, that neither the scalar nor the tensor spectrum directly probe the inflationary Hubble rate in the model under consideration: observable primordial gravitational waves are possible even for a relatively low-scale inflation. 

\subsection{Vectors}

Just like the tensor and scalar modes, the gaugid's vector perturbations do also get excited during the quasi-de Sitter phase. Similarly to the parity-even scalar mode, these obey the exact same dynamics as their counterparts in a solid---the solid's transverse phonons---leading to a vector-to-scalar ratio that scales as $(c_{T}/c_{V})^5$ with the vector speed of sound $c_{V}$ \cite{Endlich:2012pz}. Nevertheless, vector perturbations decay during the standard post-inflationary evolution of the universe, and are therefore less likely to affect the CMB in any observable way. 
For completeness, we sketch a derivation of the vector action \eqref{vaction} in gaugid inflation in Appendix \ref{vectors}.
\section{Reheating and matching the correlation functions}
\label{reheating}

\textit{After} inflation, the universe transitions into a conventional Big Bang phase, well-described by a pefect fluid with negligible anisotropic stress. Weinberg's theorem thus becomes operative in this post-inflationary epoch, and the two curvature perturbations asymptote, as usual, to the same (constant) value at large wavelengths.~However, which one, if any, of the two correlators $\langle \zeta\zeta \rangle$ and $\langle \mathcal{R} \mathcal{R} \rangle$ evaluated by the end of inflation matches onto this asymptotic value is in general dependent on the particular model of reheating. In addressing this question, our discussion will parallel that of ref.~\cite{Endlich:2012pz}, presented in the context of solid inflation. 

\vskip 0.15cm

For the rest of the argument, we will assume that reheating---that is, the transition from the inflationary phase to the standard perfect fluid-dominated era---happens instantaneously,~i.e. at most within a Hubble time interval. Moreover, we will assume that there exists a \textit{smooth} reheating surface---i.e.~a spacelike hypersurface $\Sigma_r$ separating the inflationary and Big Bang phases. What we mean by ``smooth'' here is that a coordinate system can be found on $\Sigma_r$, such that all physical quantities including the metric, as well as their first and second (space-) derivatives \textit{along} $\Sigma_r$ are finite.~Then, one can argue that the metric and its first space- and time-derivatives, as well as the $T^0_0$ (energy density) and $T^0_i$ (momentum density) components of the total matter stress tensor, evolve continuously across $\Sigma_r$, while $T^i_j$ (pressure and the various stresses) may experience a jump.~This is because $T^0_0$ and $T^0_i$ enter into the Hamiltonian and momentum \textit{constraint equations} of general relativity, which do not contain second or higher time derivatives of any quantity and are thus merely constraints on the boundary data.~(The constraint equations do contain second \textit{space} derivatives, however, and hence our definition of a ``smooth'' reheating surface.) In contrast, $T^i_j$ enter into the truly \textit{dynamical} part of Einstein's equations, meaning that this quantities may jump in going through reheating.  
From now on, we will exclusively focus on  linearized scalar perturbations, so that the two continuous quantities are the energy density $\rho$ and the scalar potential for momentum density $T^0_i = \p_i \delta q$. (For the gaugid, these are computed in appendix \ref{strten}.) 

\vskip 0.15cm


Specifying a theory of reheating requires specifying, in the first place, the `clock' that controls the end of inflation/the onset of the standard Big Bang phase. In single-field, slow roll inflation, the inflaton itself is the order parameter that provides such a clock, and the reheating surface coincides, at sufficiently large distances, with both the uniform density ($\delta\rho = 0$) and the comoving ($\delta q = 0$) hypersurfaces. 
In gaugid inflation, the most obvious candidate for the clock controlling the onset of reheating is $X$,  the only invariant in \eqref{invariants} that depends on time when evaluated on the background. 
In SFSG, and to the linear order in all (including metric) perturbations, this quantity reads
\be
\label{xpert}
X = \frac{24}{a^4}\(1+\frac{2}{3}\p^2\tilde T\)~,
\ee
where for notational convenience we have chosen to work with $\tilde T \equiv T/\sqrt{-\p^2}$. 

\vskip 0.15cm

It is not difficult to convince oneself that with this choice of $X$ as the ``clock", the reheating surface coincides with the surface of uniform energy density (but not with the comoving one). In order to see this, note that, apart from $X$, none of the invariants in \eqref{invariants} fluctuate at the linear order in the fields. Indeed, by their very construction, these invariants are insensitive to a constant rescaling of the magnetic fields, 
\be
\label{brescaling}
\delta B^I = \lambda B^I~.
\ee
To the leading order in $\lambda$ and in the fields, this transformation is realized on the parity-even scalar as $\delta\tilde T =\lambda \vec x^2/2$, which makes it clear that no quantity invariant under \eqref{brescaling} can receive a contribution linear in $\tilde T$.\footnote{Again, $\alpha$ can not contribute due to its quantum numbers under parity.} Moreover, one can show that since we are dealing with a purely magnetic background (with only the spatial components of the field strengths having non-zero expectation values), no metric perturbation can contribute to these invariants in SFSG at the linear order either.\footnote{This is because the spatial metric is not perturbed by any scalar quantity in SFSG.} Finally, being time-independent when evaluated on the background, none of these invariants transform, at the linear order in field fluctuations, under the time diff 
\be
\label{ugtimediff}
t' = t - \frac{1}{6 H}\p^2 \tilde T + \mathcal{O}\(\p^2 \tilde T\p^2 \tilde T\).
\ee
that takes us from SFSG to the gauge characterized by an unperturbed  $X$.~Since none of the building blocks of our theory are perturbed in the latter gauge, the energy density is unperturbed as well: the reheating hypersurface coincides with that of uniform energy density.

\vskip 0.15cm

One can directly check the latter statement by applying the time diff \eqref{ugtimediff} to the SFSG expression for $\delta\rho$ that we provide in eq.~\eqref{perts1} in appendix~\ref{strten} .~Applying the same diff to the momentum potential $\delta q$, on the other hand, does \textit{not} eliminate this quantity \cite{Endlich:2012pz}, meaning in particular that the reheating and comoving hypersurfaces do not coincide. We thus arrive at a conclusion that the curvature perturbation on uniform-density hypersurfaces $\zeta$ goes through reheating continuously, while the comoving curvature perturbation $\mathcal{R}$ generally experiences a jump.

\vskip 0.15cm

With our assumption that inflation terminates when $X$ drops down to a given value $X_f$, reheating is most naturally described in the gauge where $X$ does not fluctuate and inflation  ends everywhere at the same time $t'_f$ defined in eq.~\eqref{ugtimediff}.~Assuming reheating is instantaneous, the various correlators can then be computed at this time and used as the initial data for further, post-inflationary evolution. But since the difference between $t'$ and the SFSG time $t$ is at least linear in the field fluctuations,  the $\langle \zeta(t', \vec k_1) \zeta(t', \vec k_2)\rangle$ two-point function only differs from the $\langle \zeta( t, \vec k_1) \zeta( t, \vec k_2)\rangle$ correlator computed in eq.~\eqref{scalarspec} by terms of higher order in the perturbative expansion in the fields \cite{Endlich:2012pz}. 

\vskip 0.15cm

Of course, one could immagine more exotic reheating mechanisms in which the  reheating surface ends up being characterized not by a uniform density surface $\delta \rho=0$ but, more generally, by the vanishing of some linear combination,\footnote{While sufficiently general for our purposes, this assumption is of course not the most general one: the reheating surface does not necessarily have to be related to the uniform-density and/or comoving surfaces.}
\be
x\delta\rho + y H\delta q = 0, 
\ee
where $x$ and $y$ are some dimensionless constants.~From the definition \eqref{zetardef} of the two curvature perturbations it is clear that in the most general situation, both $\zeta$ and $\mathcal{R}$ experience jumps at reheating. This is because the expressions for these quantities involve the backrgound pressure, which changes discontinuously, as argued at the beginning of this section.~However, the combination 
\be
3x \zeta + y\mathcal{R} = (3x + y) C
\ee
is solely determined by the continuous metric fluctuation $C$, so that it has the same value on both sides of the reheating surface. Denoting the two gauge-invariant curvature perturbations right before (after) reheating by $\zeta_f$ and $\mathcal{R}_f$ ($\zeta_{BB}$ and $\mathcal{R}_{BB} = \zeta_{BB}$), we have 
\be
\zeta_{BB} = \frac{3x\zeta_f +y \mathcal{R}_f}{3x + y}~.
\ee
One can see, in particular, that the measured scalar perturbation $\zeta_{BB}$ is given by $\zeta_f$ ($\mathcal{R}_f$) if the reheating surface coincides with the uniform-density (comoving) hypersurface.

\vskip 0.15cm

Fast/instantaneous transition from inflation to the Big Bang phase guarantees that the initial velocity \eqref{zetadot} of $\zeta$ at the end of inflation will only result in a small relative change, $\delta \zeta/\zeta \sim \epsilon$, before this variable reaches its asymptotic value \cite{Endlich:2012pz}.~The same applies to the tensors: the graviton's wavefunction can vary at most by order $\mathcal{O}(\epsilon)$ through reheating, if it completes within a single Hubble time. Notice, though, that depending on the details of the transition, the tensor modes can be additionally sourced via a conversion of the gaugid's helicity-2 fluctuations into the metric fluctuations right after inflation terminates.~As far as reheating is fast, however, it is reasonable to assume that this effect is suppressed as well.

\vskip 0.15cm

To summarize: even though the scalar perturbations are not adiabatic in the usual sense during inflation, they get converted into the standard adiabatic perturbations shortly afterwards. Moreover, the asymptotic spectrum of adiabatic perturbations is set, under certain mild assumptions about reheating, by the $\langle \zeta\zeta \rangle$ two-point function evaluated at the end of inflation. The same applies to the primordial gravitational waves: provided that reheating completes within a single Hubble time, their asymptotic spectrum is reproduced to a very good approximation by its expression \eqref{ttspectrumtm} evaluated at the end of inflation.

\section{Summary and discussion}
\label{conclusions}

Primordial gravitational waves are among a few observables that, if detected, would provide us with unique information about the dynamics of the early universe. Given the richness of the future experimental program, there are good reasons to {hope that they will be observed in a not-so-distant future.
What makes the inflationary tensor spectrum especially interesting is its remarkable robustness \cite{Creminelli:2014wna}: the absolute majority of models predict one and the same standard expression \eqref{stdspec} for the amplitude of $\Delta^2_t$, and any deviation from that would point towards rather unconventional physics taking place in the early universe. A basic question then is: under what conditions can the spectrum of primordial gravity waves be
modified?~To our knowledge previous proposals  have been based on bremsstrahlung from a repeated production of scalar particles \cite{Senatore:2011sp,Mirbabayi:2014jqa}, rescattering of gauge fields produced by a rolling axion \cite{Sorbo:2011rz,Anber:2012du}, parity-breaking dynamics of non-abelian gauge theories (with or without axion fields) \cite{Maleknejad:2011jw,Adshead:2012kp}, and late time oscillations from a massive graviton field displaced from its minimum \cite{Dubovsky:2005dw}.

\vskip 0.15cm

In this paper we have proposed a novel and simple theory of inflation with a 
non-standard phenomenology for  both the primordial density fluctuations and the primordial gravitational waves. The model features a triplet of abelian gauge fields
in the purely magnetic background configuration shown in eq.~\eqref{op}.  This  entails a peculiar  breaking pattern of spacetime and internal symmetries, which nonetheless preserves a residual  spacial $ISO(3)$. The resulting system, which we dubbed  {\it{magnetic gaugid}},  generalizes the effective field theory of a homogeneous isotropic solid. While the small flutuations around the solid are described by three independent phonon modes of helicity 0 and 1, the gaugid counts six such modes of helicity 0, 1 and 2. The presence of a helicity-2 phonon is the main
 novelty of the gaugid.
 When considering dynamical gravity, one must consider the two additional polarizations of the graviton. Therefore small fluctuations around gaugid driven cosmology count eight modes. The dynamics of these modes is most conveniently understood by working in ``unitary gauge'', in which by using diffs and $U(1)^3$ gauge transformations one eliminates as many degrees of freedom as possible from the gauge fields. The modes living inside the gauge field then consist of
 a pseudoscalar $\alpha$ and a transverse traceless tensor $E_{ij}$. The remaining degrees of freedom appear in the traceless (but not transverse) spatial metric $\gamma_{ij}$. Classifying the finite momentum modes according to their helicity $h$, one has  $h_\alpha=0$ and $h_E=\pm 2$. On the other hand, $\gamma_{ij}$
 decomposes as $0\oplus1\oplus 2$ for a total of 5 polarizations. The scalar and vector modes living inside $\gamma_{ij}$ are fully analogous to their counterparts in solid inflation. In particular one can reasonably argue that the scalar mode $T$ converts at reheating into the standard density perturbation $\zeta$. The main novelty with respect to solid inflation is represented by the additional tensor mode $E_{ij}$.

\vskip 0.15cm 
 
  As it turns out, following ref.~\cite{Weinberg:2003sw}, a good deal about the dynamics of fluctuations can be grasped by considering observability of the various modes when stretched outside the horizon. Observability is established  by considering the possible residual global gauge symmetries (diffeomorphisms included). In the present case it turns out that  all the zero momentum modes of $\gamma_{ij}$ and of the magnetic fields associated to $\alpha$ and $E_{ij}$ transform under residual shifts, with the exception of the combination $\tilde \gamma_{ij}$ defined in eq.~\eqref{anisotropy}. The physical combination $\tilde \gamma_{ij}$ has therefore a non-trivial dynamics  outside the horizon.
This  is associated to a slow-roll suppressed positive graviton mass squared $m_\gamma^2\propto \epsilon$ which  slowly drives $\tilde\gamma_{ij}$ to zero like $a^{-\epsilon}$. In other words, outside the horizon 
\beq
\label{limit}
\gamma_{ij}\to  \frac{\epsilon_{ikl}\p_k E_{lj}+\epsilon_{jkl}\p_k E_{li}}{2}
\eeq
over a number of e-foldings $N\gsim 1/\epsilon$. At this point the dynamics that leads to the result in eq.~\eqref{r} for primordial gravity waves is qualitatively rather clear. At subhorizon distances, $E_{ij}$ coincides with the genuine helicity 2-phonon of the gaugid. Its Bunch-Davies quantum fluctuations are therefore larger than those of gravity,
 corresponding to the scale  of the gaugid being slow roll suppressed. As one can easily deduce from the $E_{ij}$ kinetic term in eq.~\eqref{actiontt},  one schematically  has $\langle E E\rangle \propto 1/\epsilon$.  Once outside the horizon, by eq.~\eqref{limit},
 these large $E_{ij}$ fluctuations are slowly, but implacably, transferred to $\gamma_{ij}$.
Because of that, one can in principle even obtain $\langle \gamma\gamma\rangle\propto 1/\epsilon$, corresponding to a tensor-scalar ratio $r\sim 1$. That is of course already ruled out, but the model possesses a sufficient number of free parameters to easily give an acceptable but sizeable $r$,  independently of the scale of inflation! This is the main result of our paper.

\vskip 0.15cm 
 
It should be made clear that, as long as they are outside the horizon, the large metric fluctuations we have just discussed are not truly observable, given that they correspond to a suppressed $\tilde \gamma_{ij}$. But that is precisely like for the  adiabatic modes in single field inflation: the large $\gamma_{ij}$
becomes observable only after re-entering the horizon during the standard Big Bang cosmological phase. Notice also that, very much like for solid inflation, and unlike standard inflation, our model must rely on assumptions about the process of reheating in order to make definite predictions. On general grounds we can picture
 reheating as a phase transition era where the matter sector transits from a gaugid phase to an ordinary cosmological thermal fluid phase. At the phase transition the degrees of freedom in the matter sector behave discontinuously, in particular, $E_{ij}$ disappears. It is however reasonable to expect the gravitational modes to behave more regularly in such a way that
 $\gamma_{ij}$, after reheating, will only change by $O(\epsilon)$ compared to its value just before. In section~\ref{reheating}, following ref.~\cite{Endlich:2012pz}, we have indeed argued that that must be  the case, provided reheating happens fast, within a Hubble time.


\vskip 0.15cm

Our work can be extended in several directions. It would be interesting to study the structure of higher point functions.~Those involving tensor
modes, e.g.~$\langle \gamma_{ij}\zeta \zeta\rangle$, will definitely differ from all other scenarios of inflation, including solid inflation. 
As concerns the $\langle \zeta\zeta \zeta\rangle$ three point function, we  expect
a result similar to that of solid inflation.~This is because, as we have discussed in section~\ref{scalarsec}, the dynamics of gaugids closely resembles that of solids as far as one truncates to the parity-even scalar sector consisting solely of the field $T$. There certainly are additional effects associated to extra pseudoscalar fields present in gaugid inflation: $\alpha$ and the temporal components of the gauge fields $a_{i0} = \p_i\chi/\sqrt{-\p^2}$; however, due to the invariance of the perturbation lagrangian under parity, these effects can only kick in for $n$-point functions higher than cubic, or at higher loop level. One could also consider how things change by allowing parity breaking into the game, in such a way that there exists
both a magnetic and an electric background field. In that case the additional scalar $\alpha$, which is un-mixed because of parity in the magentic gaugid, would definitely play a role in the dynamics of $\zeta$.}

\subsection*{Acknowledgements}
We would like to thank Jose Beltr\'an Jim\'enez, Paolo Creminelli, Sergei Dubovsky, Sasha Monin, Joao Penedones, Sergey Sibiryakov, Enrico Trincherini and especially Alberto Nicolis for valuable discussions. F.P. acknowledges 
A*MIDEX project (n ANR-11-IDEX-0001-02) funded by the  "Investissements d'Avenir" French Government program, managed by the French National Research Agency (ANR). The work of D.P. and R.R. is partially supported by the Swiss National Science Foundation  under contracts 200020-150060 and 200020-169696.

\appendix

\section{The perturbation Lagrangian}
\label{appa}

In this appendix we discuss, in some more detail, the calculation of the quadratic perturbation Lagrangian in the gaugid phase.~We start with exploring the sub-horizon regime, where mixing with gravity becomes irrelevant. This is followed by a discussion of the full calculation, incorporating dynamical gravity, of the scalar and vector fluctuations' quadratic action.

\subsection{The flat space/subhorizon fluctuations}
\label{appa1}

We start off by studying the spectrum of fluctuations in the theory \eqref{themodel}.
We work in the $U(1)^3$ gauge where $S=S_i=0$, so that the field perturbations can be decomposed into the various helicity modes as follows
\be
\label{helicitydec}
a_{ij} = \alpha\delta_{ij} + E_{ij} + \epsilon_{ijk}\(\frac{\partial_k T}{\sqrt{-\p^2}} + V_k\)~,\qquad  a_{i0} = \frac{\partial_i \chi}{\sqrt{-\p^2}} + B_i~,
\ee
where $E_{ij}$ is a symmetric, transverse and traceless tensor, while $T_i$ and $B_i$ are transverse vectors.

\vskip 0.15cm

Zooming onto distance/time scales much shorter than the Hubble horizon, the expansion of the universe becomes irrelevant and \eqref{themodel} reduces to the following expression at the quadratic order in perturbations
\beal
\label{masteract}
S^{(2)} =~& \int \dd^4 x \bigg[-\frac{XP_X}{24}f_i^{~\mu\nu}f_{i\mu\nu} + \frac{X P_{X}}{72}(\epsilon_{ijk}f_{ikj})^2\\ &-M^4_1\(f_{ijk}f_{ijk}+f_{ijk}f_{jik}-\frac{3}{2}f_{iik}f_{jjk}\)+M^4_2\(f_{ij0}f_{ij0}+f_{ij0}f_{ji0}\)\bigg]~,
\eeal
where we have set the scale factor to one and have neglected mixing with metric perturbations. Since we are ultimately interested in quasi de-Sitter perturbations, we have also used the relation \eqref{quasidscond}. 
A straightforward expansion of the above action into the helicity modes yields
\begin{align}
S^{(2)} & =\int~ \dd^4 x \bigg[\(\frac{XP_X}{4}  +6M^4_2\)\dot{\alpha}^2-\frac{XP_X}{6}  (\p\alpha)^2-\(\frac{XP_X}{6}  +4M^4_2\)\frac{\p^2\chi}{\sqrt{-\p^2}}~\dot{\alpha} \\ \nonumber 
& - \(\frac{XP_X}{12}+2M^4_2\)\chi\p^2\chi+\frac{XP_X}{6}~ \dot{T}^2 - \(2M^4_1-\frac{XP_X}{18}\)(\p T)^2+\frac{XP_X}{6}~ \dot{V_k}^2 \\ \nonumber
&-\(\frac{XP_X}{12}+\frac{3}{2}M^4_1\)(\p V_k)^2-\frac{XP_X}{6}~\epsilon_{ijk}\p_j B_i \dot V_k +\(\frac{XP_X}{12}+M^4_2\)(\p B_k)^2\\ \nonumber
&+\(\frac{XP_X}{12}+2M^4_2\)\dot E^2_{ij}-\(\frac{XP_X}{12}+3M^4_1\)(\p E_{ij})^2\bigg]~.
\end{align}
As before, $\chi$ and $B_i$ are non-dynamical fields, and can be integrated out through their respective equations of motion:
\beq
\chi= -\frac{\dot \alpha}{\sqrt{-\p^2}}~,\qquad B_i = \frac{XP_X}{XP_X+12M^4_2}\epsilon_{ijk}\p^{-2}\p_j\dot V_k~.
\eeq
Plugging the latter solutions back into the action then yields the final expression for the dynamical modes' quadratic Lagrangian
\begin{align}
\label{lagflat}
S^{(2)} & =\int~ \dd^4 x \bigg[\(\frac{XP_X}{6}  +4M^4_2\)\dot{\alpha}^2-\frac{XP_X}{6}  (\p\alpha)^2+\frac{XP_X}{6}~ \dot{T}^2 - \(2M^4_1-\frac{XP_X}{18}\)(\p T)^2 \\ \nonumber
&+\frac{X P_X \(X P_X+24 M^4_2\)}{12 (X P_X+12M^4_2)}~ \dot{V_k}^2-\(\frac{XP_X}{12}+\frac{3}{2}M^4_1\)(\p V_k)^2+\(\frac{XP_X}{12}+2M^4_2\)\dot E^2_{ij}\\ \nonumber &-\(\frac{XP_X}{12}+3M^4_1\)(\p E_{ij})^2\bigg]~.
\end{align}
One can straightforwardly show, that all modes are stable (free from both the ghost and gradient instabilities) and subluminal for the choice of the parameters satisfying eq.~\eqref{stcond}.

\subsection{Scalar perturbations including mixing with gravity}
\label{scalarmixing}

A brute-force expansion of the action \eqref{themodel} to the quadratic order in the scalars yields 
\begin{align}
S^{(2)} & =\int~ \dd^4 x ~a^3\bigg[\(\frac{XP_X}{4}  +6M^4_2\)a^2\dot{\alpha}^2-\frac{XP_X}{6}  (\p\alpha)^2+\(\frac{XP_X}{6}  +4M^4_2\)a^2 \dot{\alpha} \sqrt{-\p^2}\chi\\ \nonumber 
& - \(\frac{XP_X}{12}+2M^4_2\)a^2\chi\p^2\chi+\frac{XP_X}{6}~a^2 \dot{T}^2 - \(2M^4_1-\frac{XP_X}{18}\)(\p T)^2 - 3\mpl^2 H^2\delta N^2 \\ \nonumber 
&+\frac{2XP_X}{3}\delta N \sqrt{-\p^2}T - 2\mpl^2 H\delta N ~\frac{1}{a^2}\p^2\psi-\frac{2XP_X}{3}\psi\sqrt{-\p^2} \dot T+\frac{2XP_X}{3} \frac{1}{a^2}(\p\psi)^2 \bigg]~,
\end{align}
where we have denoted $a_{i0}=\p_i\chi/\sqrt{-\p^2}$.
One can see that the action splits into two independent sectors, featuring fields with opposite parity. Integrating out the non-dynamical $\chi$ results in the simple Lagrangian for the gaugid's parity-odd scalar fluctuation in eq.~\eqref{alphaaction}. 

\vskip 0.15cm

The parity-even sector, on the other hand, features two non-dynamical degrees of freedom -- the lapse and (scalar) shift perturbations. Integrating these out, see eq.~\eqref{lapseandshift}, one arrives at the final action \eqref{Taction} for the gaugid's $T$ mode.

\subsection{Vector perturbations including mixing with gravity}

\label{vectors}

Expanding the action \eqref{themodel} on a quasi-de Sitter background around the magnetic gaugid configuration, one deals with three vector degrees of freedom: the transverse shift perturbation for which we use the following notation
\beal
N^i= N^T_i~,
\eeal  
and the helicity-1 collective modes $B_k$ and $V_k$, defined in Eq. \eqref{helicitydec}. The former two are non-dynamical, which can be seen from the explicit form of the quadratic Lagrangian for vector fluctuations
\begin{align}
S^{(2)} & =\int~ \dd^4 x ~a^3\bigg[ \frac{1}{4}\mpl^2 (\p N^T_k)^2+\frac{2}{3} a^2XP_X N^T_kN^T_k-\frac{2}{3} a^2XP_X\dot V_k N^T_k  \nn \\&-\frac{1}{3} a^2XP_X\epsilon_{ijk}\p_kB_i N^T_j
+\frac{1}{6} a^2XP_X\dot V_k^2-\(\frac{1}{12}XP_X+\frac{3}{2}M_1^4\)(\p V_k)^2\nn \\&-\frac{1}{6}a^2XP_X\epsilon_{ijk}\p_jB_i\dot V_k+a^2 \(\frac{1}{12}XP_X+M_2^4\)(\p B_k)^2\bigg]~.
\end{align}
Integrating out the non-propagating fields then generates the (momentum-space) action for the only dynamical vector mode, given in eq.~\eqref{vaction}.

\vskip 0.15cm

This action exactly coincides with the one describing the transverse phonons in solid inflation, so all results of ref.~\cite{Endlich:2012pz} concerning the vector perturbations carry over unaltered to the model under consideration.

\section{Quantum mechanics of multiple fields on de Sitter}
\label{gws}

In this appendix, we fill in the gaps in our derivation of the expression for the spectrum of the primordial gravitational waves \eqref{ttspectrumtm}.

\vskip 0.15cm

To start with, consider a set of $N$ fields $\phi_\alpha$, with the following quadratic action in momentum space
\begin{equation}
\label{multiplefields}
S = \frac12 \int d\tau \frac{d^3k}{(2\pi)^3} \left[ \phi'_\alpha(\tau,\vec k) \phi'_\alpha(\tau,\text{-}\vec k) - k^2 \phi_\alpha(\tau,\vec k) M_{\alpha\beta}(k, \tau) \phi_\beta(\tau,\text{-}\vec k)\right]\, ,
\end{equation}
where the fields are canonically normalized on (quasi-) de Sitter spacetime and $M_{\alpha\beta}$ is some real and symmetric matrix.~We will assume $\phi_\alpha$ stem from real position-space fields, so that $\phi_\alpha^\dagger(\tau,\vec k) = \phi_\alpha(\tau,-\vec k)$.

\vskip 0.15cm

The equations of motion that follow from the above action,
\begin{equation} \label{evoluzione}
 \frac{d^2 \phi_\alpha(\tau,\vec k)}{d \tau^2} + k^2 M_{\alpha\beta}(k, \tau) \phi_\beta(\tau,\vec k) = 0\, ,
\end{equation}
generically admit $2N$ independent solutions, $f_\alpha^{(n)}(t, k)$ and $f_\alpha^{(n)}(t, k)^*$, where $n = 1\dots N$ labels the $N$ modes, present in the theory. Choosing a vacuum corresponds to choosing the mode functions $f_\alpha^{(n)}(t, k)$, with respect to which the fields are quantized:\footnote{Notice that the isotropy of the background constrains $f_\alpha^{(n)}$ to only depend on the \textit{magnitude} of $\vec k$, but not on the direction.}
 \begin{equation} \label{modedecomposition}
 \phi_n(\tau, \vec k) =  f_\alpha^{(n)}(\tau,k)~ a_n (\vec k) + f_\alpha^{(n)}(\tau,k)^* ~a_n^\dagger(\text{-}\vec k)\, 
 \end{equation}
(summation over the repeated index $n$ is assumed).~Here,  $a_n(\vec k)$ ($a^\dagger_n(\vec k)$) are the standard time-independent annihilation (creation) operators for the $N$ modes. 
The (momentum-space) canonical commutation relations read 
\begin{align}
\label{cancom}
\begin{split}
[\phi_\alpha(\tau,\vec k),\phi'_\beta(\tau,\vec q) ] &= (2\pi)^3 i \delta^{(3)}(\vec k+\vec q)\delta_{\alpha\beta},~\\
[\phi_\alpha(\tau,\vec k),\phi_\beta(\tau,\vec q) ] &= 0~,\\
[\phi'_\alpha(\tau,\vec k),\phi'_\beta(\tau,\vec q) ] &= 0~,
\end{split}
\end{align}
with all other commutators obtained from these using the relation $\phi_\alpha^\dagger(\tau,\vec k) = \phi_\alpha(\tau,-\vec k)$. 

\vskip 0.15cm

Furthermore, we will assume that the mode functions satisfy the following orthonormality conditions:
\begin{align}
\label{wronskian}
\begin{split}
\mathcal{W}\(f^{(m)}(\tau,k), f^{(n)}(\tau,k)\) &\equiv f_\alpha^{(m)}(\tau,k) ~\frac{d{{f_\alpha^{(n)}}(\tau,k)^{*}}}{d\tau}- \frac{d{f_\alpha^{(m)}}(\tau,k)}{d\tau} {f^{(n)}_\alpha}(\tau,k)^* = i \delta_{mn}~, \\ \mathcal{W}\(f^{(m)}(\tau,k)^*, f^{(n)}(\tau,k)\) &=\mathcal{W}\(f^{(m)}(\tau,k), f^{(n)}(\tau,k)^*\) =  0~.
\end{split}
\end{align}
The \textit{Wronskian} $\mathcal{W}\(f(\tau,k),g(\tau,k)\)$ can be easily shown to be time-independent for any two solutions $f$ and $g$ of eq.~\eqref{evoluzione}.
This means, in particular, that any set of mode functions satisfying eq.~\eqref{wronskian} at some reference time will satisfy the same set of relations at \textit{any other time}.

\vskip 0.15cm

The relation \eqref{modedecomposition} between the field and creation/annihilation operators can be inverted with the help of eq.~\eqref{wronskian}
\begin{align}
a_n(\vec k) &= - i \mathcal{W}\(\phi(\tau,\vec k), f^{(n)} (\tau,k)\) ~, \\
a^\dagger_n(\vec k) &=  i \mathcal{W}\(\phi^\dagger (\tau,\vec k), f^{(n)} (\tau,k)^*\)~.
\end{align}
These expressions, together with the canonical commutation relations \eqref{cancom}, yield the standard algebra for the creation and annihilation operators
\begin{align}
\label{cancom1}
\begin{split}
[a_m (\vec k),a^\dagger_n(\vec q)]&=- i (2\pi)^3 \delta^{(3)}(\vec{k}-\vec{q})~\mathcal{W}\(f^{(m)}(\tau,k), f^{(n)}(\tau,k)\) = (2\pi)^3 \delta^{(3)}(\vec{k}-\vec{q})\delta_{mn}~,\\
[a_m (\vec k),a_n(\vec q)]&=- i (2\pi)^3 \delta^{(3)}(\vec{k}+\vec{q})~\mathcal{W}\(f^{(m)}(\tau,k)^*, f^{(n)}(\tau,k)\)=0~,\\
[a^\dagger_m (\vec k),a^\dagger_n(\vec q)]&=- i (2\pi)^3 \delta^{(3)}(\vec{k}+\vec{q})~\mathcal{W}\(f^{(m)}(\tau,k), f^{(n)}(\tau,k)^\star\)=0~,
\end{split}
\end{align}
where we have used the orthonormality of the mode functions \eqref{wronskian}. 

\vskip 0.15cm

To summarize, provided one can find a set of mode functions such that the orthonormality conditions \eqref{wronskian} are satisfied at some reference time (say at $\tau\to -\infty$),
one is guaranteed to have creation and annihilation opreators obeying the standard algebra for \textit{all} $\tau$.

\vskip 0.15cm

In most physical situations, the structure of the mixing terms in \eqref{multiplefields} is such that the off-diagonal components in $M$ decay sufficiently fast and the action decouples into $N$ independent harmonic oscillators at short wavelengths:
\begin{equation} \label{diagonal}
M \ \overset{k\tau \rightarrow -\infty}\longrightarrow  \ \text{diag}\{c_1^2,\ c_2^2,\ \dots\, , \ c_N^2\}~.
\end{equation}
In this case, the physical, \textit{Bunch-Davies} vacuum corresponds to the following choice of the mode functions 
\begin{equation} \label{choice}
f_\alpha^{(n)}(\tau,k) \ \overset{k\tau \rightarrow -\infty}\longrightarrow \ \delta_\alpha^n \ \frac{e^{- ik\int  c_n(\tau)  d\tau}}{\ \sqrt{2 c_n(\tau) k}\ }\, ~,
\end{equation}
where we have allowed for (adiabatically) time-dependent speeds of propagation for various modes.
It is easy to see that these mode functions do indeed satisfy the orthonormality conditions~\eqref{wronskian}. Note, however, that under some special circumstances the off-diagonal components in $M$ may not decay fast enough for $k\tau\to -\infty$~, in which case the asymptotic Bunch-Davies mode functions will have a form, different from \eqref{choice}. 

\section{The perturbed stress tensor}
\label{strten}

Varying the action \eqref{themodel} with respect to the metric, one obtains the following expression for the gaugid's stress-energy tensor
\beal
\label{stresstensor}
T^{\mu}_{\nu} = -\delta^{\mu}_{\nu}\bigg[P+\(27 M_1^4+18 M_2^4\) I_2-72 M_2^4 W\bigg]+ 4 P_X F_{I~~\alpha}^{~\mu} F_{I\nu}^{~~\alpha}~.
\eeal
The perturbed scalar quantities are then defined as follows
\beal
T^0_0&=-(\rho+\delta\rho)~,  \\
T^0_i&=\p_i\delta q~, \\
T^i_j&=\delta^i_j(p+\delta p)+\sigma^i_{j}~,
\eeal
were $\sigma^i_j$ is the (scalar) anisotropic stress ($\sigma^i_i=0$). 
A straightforward calculation yields
\beal
\label{perts1}
\delta\rho &=-\mpl^2 H^2\epsilon~\sqrt{-\p^2}T~,  \\
\delta q &= \mpl^2 H^2 \epsilon ~\(2\psi - a^2 \frac{\dot T}{\sqrt{-\p^2}}\)~,  \\
\sigma^i_j&=\mpl^2 H^2 \epsilon~\(\p_i\p_j-\frac{1}{3}\delta_{ij}\p^2\)\frac{T}{\sqrt{-\p^2}}~.
\eeal
Notice that by using $3M_P^2H^2=\rho$ and $\epsilon=-\dot H/H^2$ we can write
\beq
\rho+\delta \rho=3M_P^2\left(1-\frac{\sqrt{-\partial^2} T}{6H}\frac{d}{dt}\right) H^2
\eeq
from which one can immediately check that the time diff in eq.~\eqref{ugtimediff} sets $\delta\rho=0$.
In the spatially flat gauge, the comoving curvature perturbation $\mathcal{R}$ and the curvature perturbation on uniform-density surfaces $\zeta$ read 
\beal
\mathcal{R} &= - \frac{H}{\rho+p}\delta q~,  \\
\zeta &=\frac{\delta\rho}{3(\rho+p)}~.
\eeal   
Expressing the lapse and shift perturbations in terms of $T$ via the Hamiltonian and momentum constraints, see eq.~\eqref{lapseandshift}, and using eqs.~\eqref{stresstensor} through \eqref{perts1}, one obtains the expressions for $\mathcal{R}$ and $\zeta$ given in eq.~\eqref{randz}.
Our definitions for the two curvature perturbations are related to those of Ref. \cite{Endlich:2012pz} through a flip of sign for both of these quantities.


\bibliographystyle{utphys}
\addcontentsline{toc}{section}{References}
\bibliography{bibliography}

\providecommand{\href}[2]{#2}\begingroup\raggedright\begin{thebibliography}{10}

\bibitem{Nicolis:2015sra}
A.~Nicolis, R.~Penco, F.~Piazza, and R.~Rattazzi, ``{Zoology of condensed
  matter: Framids, ordinary stuff, extra-ordinary stuff},''
  \href{http://dx.doi.org/10.1007/JHEP06(2015)155}{{\em JHEP} {\bfseries 06}
  (2015) 155},
\href{http://arxiv.org/abs/1501.03845}{{\ttfamily arXiv:1501.03845 [hep-th]}}.

\bibitem{Creminelli:2006xe}
P.~Creminelli, M.~A. Luty, A.~Nicolis, and L.~Senatore, ``{Starting the
  Universe: Stable Violation of the Null Energy Condition and Non-standard
  Cosmologies},'' \href{http://dx.doi.org/10.1088/1126-6708/2006/12/080}{{\em
  JHEP} {\bfseries 0612} (2006) 080},
\href{http://arxiv.org/abs/hep-th/0606090}{{\ttfamily arXiv:hep-th/0606090
  [hep-th]}}.

\bibitem{Cheung:2007st}
C.~Cheung, P.~Creminelli, A.~L. Fitzpatrick, J.~Kaplan, and L.~Senatore, ``{The
  Effective Field Theory of Inflation},''
  \href{http://dx.doi.org/10.1088/1126-6708/2008/03/014}{{\em JHEP} {\bfseries
  0803} (2008) 014},
\href{http://arxiv.org/abs/0709.0293}{{\ttfamily arXiv:0709.0293 [hep-th]}}.

\bibitem{Maldacena:2002vr}
J.~M. Maldacena, ``{Non-Gaussian features of primordial fluctuations in single
  field inflationary models},''
  \href{http://dx.doi.org/10.1088/1126-6708/2003/05/013}{{\em JHEP} {\bfseries
  0305} (2003) 013},
\href{http://arxiv.org/abs/astro-ph/0210603}{{\ttfamily arXiv:astro-ph/0210603
  [astro-ph]}}.

\bibitem{Creminelli:2004yq}
P.~Creminelli and M.~Zaldarriaga, ``{Single field consistency relation for the
  3-point function},''
  \href{http://dx.doi.org/10.1088/1475-7516/2004/10/006}{{\em JCAP} {\bfseries
  0410} (2004) 006},
\href{http://arxiv.org/abs/astro-ph/0407059}{{\ttfamily arXiv:astro-ph/0407059
  [astro-ph]}}.

\bibitem{Cheung:2007sv}
C.~Cheung, A.~L. Fitzpatrick, J.~Kaplan, and L.~Senatore, ``{On the consistency
  relation of the 3-point function in single field inflation},''
  \href{http://dx.doi.org/10.1088/1475-7516/2008/02/021}{{\em JCAP} {\bfseries
  0802} (2008) 021},
\href{http://arxiv.org/abs/0709.0295}{{\ttfamily arXiv:0709.0295 [hep-th]}}.

\bibitem{Hinterbichler:2012nm}
K.~Hinterbichler, L.~Hui, and J.~Khoury, ``{Conformal Symmetries of Adiabatic
  Modes in Cosmology},''
  \href{http://dx.doi.org/10.1088/1475-7516/2012/08/017}{{\em JCAP} {\bfseries
  1208} (2012) 017},
\href{http://arxiv.org/abs/1203.6351}{{\ttfamily arXiv:1203.6351 [hep-th]}}.

\bibitem{Creminelli:2012qr}
P.~Creminelli, A.~Joyce, J.~Khoury, and M.~Simonovic, ``{Consistency Relations
  for the Conformal Mechanism},''
  \href{http://dx.doi.org/10.1088/1475-7516/2013/04/020}{{\em JCAP} {\bfseries
  1304} (2013) 020},
\href{http://arxiv.org/abs/1212.3329}{{\ttfamily arXiv:1212.3329 [hep-th]}}.

\bibitem{Hinterbichler:2013dpa}
K.~Hinterbichler, L.~Hui, and J.~Khoury, ``{An Infinite Set of Ward Identities
  for Adiabatic Modes in Cosmology},''
  \href{http://dx.doi.org/10.1088/1475-7516/2014/01/039}{{\em JCAP} {\bfseries
  1401} (2014) 039},
\href{http://arxiv.org/abs/1304.5527}{{\ttfamily arXiv:1304.5527 [hep-th]}}.

\bibitem{Bordin:2017ozj}
L.~Bordin, P.~Creminelli, M.~Mirbabayi, and J.~Noreña, ``{Solid
  Consistency},'' \href{http://dx.doi.org/10.1088/1475-7516/2017/03/004}{{\em
  JCAP} {\bfseries 1703} no.~03, (2017) 004},
\href{http://arxiv.org/abs/1701.04382}{{\ttfamily arXiv:1701.04382
  [astro-ph.CO]}}.

\bibitem{Senatore:2009gt}
L.~Senatore, K.~M. Smith, and M.~Zaldarriaga, ``{Non-Gaussianities in Single
  Field Inflation and their Optimal Limits from the WMAP 5-year Data},''
  \href{http://dx.doi.org/10.1088/1475-7516/2010/01/028}{{\em JCAP} {\bfseries
  1001} (2010) 028},
\href{http://arxiv.org/abs/0905.3746}{{\ttfamily arXiv:0905.3746
  [astro-ph.CO]}}.

\bibitem{Senatore:2010wk}
L.~Senatore and M.~Zaldarriaga, ``{The Effective Field Theory of Multifield
  Inflation},'' \href{http://dx.doi.org/10.1007/JHEP04(2012)024}{{\em JHEP}
  {\bfseries 04} (2012) 024},
\href{http://arxiv.org/abs/1009.2093}{{\ttfamily arXiv:1009.2093 [hep-th]}}.

\bibitem{Piazza:2013coa}
F.~Piazza and F.~Vernizzi, ``{Effective Field Theory of Cosmological
  Perturbations},''
  \href{http://dx.doi.org/10.1088/0264-9381/30/21/214007}{{\em Class. Quant.
  Grav.} {\bfseries 30} (2013) 214007},
\href{http://arxiv.org/abs/1307.4350}{{\ttfamily arXiv:1307.4350 [hep-th]}}.

\bibitem{Pirtskhalava:2015ebk}
D.~Pirtskhalava, L.~Santoni, and E.~Trincherini, ``{Constraints on Single-Field
  Inflation},''
\href{http://arxiv.org/abs/1511.01817}{{\ttfamily arXiv:1511.01817 [hep-th]}}.

\bibitem{Garriga:1999vw}
J.~Garriga and V.~F. Mukhanov, ``{Perturbations in k-inflation},''
  \href{http://dx.doi.org/10.1016/S0370-2693(99)00602-4}{{\em Phys.Lett.}
  {\bfseries B458} (1999) 219--225},
\href{http://arxiv.org/abs/hep-th/9904176}{{\ttfamily arXiv:hep-th/9904176
  [hep-th]}}.

\bibitem{Alishahiha:2004eh}
M.~Alishahiha, E.~Silverstein, and D.~Tong, ``{DBI in the sky},''
  \href{http://dx.doi.org/10.1103/PhysRevD.70.123505}{{\em Phys.Rev.}
  {\bfseries D70} (2004) 123505},
\href{http://arxiv.org/abs/hep-th/0404084}{{\ttfamily arXiv:hep-th/0404084
  [hep-th]}}.

\bibitem{ArkaniHamed:2003uz}
N.~Arkani-Hamed, P.~Creminelli, S.~Mukohyama, and M.~Zaldarriaga, ``{Ghost
  inflation},'' \href{http://dx.doi.org/10.1088/1475-7516/2004/04/001}{{\em
  JCAP} {\bfseries 0404} (2004) 001},
\href{http://arxiv.org/abs/hep-th/0312100}{{\ttfamily arXiv:hep-th/0312100
  [hep-th]}}.

\bibitem{Linde:1996gt}
A.~D. Linde and V.~F. Mukhanov, ``{Nongaussian isocurvature perturbations from
  inflation},'' \href{http://dx.doi.org/10.1103/PhysRevD.56.R535}{{\em
  Phys.Rev.} {\bfseries D56} (1997) 535--539},
\href{http://arxiv.org/abs/astro-ph/9610219}{{\ttfamily arXiv:astro-ph/9610219
  [astro-ph]}}.

\bibitem{Enqvist:2001zp}
K.~Enqvist and M.~S. Sloth, ``{Adiabatic CMB perturbations in pre - big bang
  string cosmology},''
  \href{http://dx.doi.org/10.1016/S0550-3213(02)00043-3}{{\em Nucl.Phys.}
  {\bfseries B626} (2002) 395--409},
\href{http://arxiv.org/abs/hep-ph/0109214}{{\ttfamily arXiv:hep-ph/0109214
  [hep-ph]}}.

\bibitem{Lyth:2001nq}
D.~H. Lyth and D.~Wands, ``{Generating the curvature perturbation without an
  inflaton},'' \href{http://dx.doi.org/10.1016/S0370-2693(01)01366-1}{{\em
  Phys.Lett.} {\bfseries B524} (2002) 5--14},
\href{http://arxiv.org/abs/hep-ph/0110002}{{\ttfamily arXiv:hep-ph/0110002
  [hep-ph]}}.

\bibitem{Dvali:2003em}
G.~Dvali, A.~Gruzinov, and M.~Zaldarriaga, ``{A new mechanism for generating
  density perturbations from inflation},''
  \href{http://dx.doi.org/10.1103/PhysRevD.69.023505}{{\em Phys.Rev.}
  {\bfseries D69} (2004) 023505},
\href{http://arxiv.org/abs/astro-ph/0303591}{{\ttfamily arXiv:astro-ph/0303591
  [astro-ph]}}.

\bibitem{Creminelli:2014wna}
P.~Creminelli, J.~Gleyzes, J.~Norena, and F.~Vernizzi, ``{Resilience of the
  standard predictions for primordial tensor modes},''
\href{http://arxiv.org/abs/1407.8439}{{\ttfamily arXiv:1407.8439
  [astro-ph.CO]}}.

\bibitem{Dubovsky:2005xd}
S.~Dubovsky, T.~Gregoire, A.~Nicolis, and R.~Rattazzi, ``{Null energy condition
  and superluminal propagation},''
  \href{http://dx.doi.org/10.1088/1126-6708/2006/03/025}{{\em JHEP} {\bfseries
  03} (2006) 025},
\href{http://arxiv.org/abs/hep-th/0512260}{{\ttfamily arXiv:hep-th/0512260
  [hep-th]}}.

\bibitem{Bucher:1998mh}
M.~Bucher and D.~N. Spergel, ``{Is the dark matter a solid?},''
  \href{http://dx.doi.org/10.1103/PhysRevD.60.043505}{{\em Phys. Rev.}
  {\bfseries D60} (1999) 043505},
\href{http://arxiv.org/abs/astro-ph/9812022}{{\ttfamily arXiv:astro-ph/9812022
  [astro-ph]}}.

\bibitem{Gruzinov:2004ty}
A.~Gruzinov, ``{Elastic inflation},''
  \href{http://dx.doi.org/10.1103/PhysRevD.70.063518}{{\em Phys. Rev.}
  {\bfseries D70} (2004) 063518},
\href{http://arxiv.org/abs/astro-ph/0404548}{{\ttfamily arXiv:astro-ph/0404548
  [astro-ph]}}.

\bibitem{Endlich:2012pz}
S.~Endlich, A.~Nicolis, and J.~Wang, ``{Solid Inflation},''
  \href{http://dx.doi.org/10.1088/1475-7516/2013/10/011}{{\em JCAP} {\bfseries
  1310} (2013) 011},
\href{http://arxiv.org/abs/1210.0569}{{\ttfamily arXiv:1210.0569 [hep-th]}}.

\bibitem{Array:2015xqh}
{\bfseries BICEP2, Keck Array} Collaboration, P.~A.~R. Ade {\em et~al.},
  ``{Improved Constraints on Cosmology and Foregrounds from BICEP2 and Keck
  Array Cosmic Microwave Background Data with Inclusion of 95 GHz Band},''
  \href{http://dx.doi.org/10.1103/PhysRevLett.116.031302}{{\em Phys. Rev.
  Lett.} {\bfseries 116} (2016) 031302},
\href{http://arxiv.org/abs/1510.09217}{{\ttfamily arXiv:1510.09217
  [astro-ph.CO]}}.

\bibitem{Maleknejad:2011jw}
A.~Maleknejad and M.~M. Sheikh-Jabbari, ``{Gauge-flation: Inflation From
  Non-Abelian Gauge Fields},''
  \href{http://dx.doi.org/10.1016/j.physletb.2013.05.001}{{\em Phys. Lett.}
  {\bfseries B723} (2013) 224--228},
\href{http://arxiv.org/abs/1102.1513}{{\ttfamily arXiv:1102.1513 [hep-ph]}}.

\bibitem{Adshead:2012kp}
P.~Adshead and M.~Wyman, ``{Chromo-Natural Inflation: Natural inflation on a
  steep potential with classical non-Abelian gauge fields},''
  \href{http://dx.doi.org/10.1103/PhysRevLett.108.261302}{{\em Phys. Rev.
  Lett.} {\bfseries 108} (2012) 261302},
\href{http://arxiv.org/abs/1202.2366}{{\ttfamily arXiv:1202.2366 [hep-th]}}.

\bibitem{Bartolo:2013msa}
N.~Bartolo, S.~Matarrese, M.~Peloso, and A.~Ricciardone, ``{Anisotropy in solid
  inflation},'' \href{http://dx.doi.org/10.1088/1475-7516/2013/08/022}{{\em
  JCAP} {\bfseries 1308} (2013) 022},
\href{http://arxiv.org/abs/1306.4160}{{\ttfamily arXiv:1306.4160
  [astro-ph.CO]}}.

\bibitem{Endlich:2013jia}
S.~Endlich, B.~Horn, A.~Nicolis, and J.~Wang, ``{Squeezed limit of the solid
  inflation three-point function},''
  \href{http://dx.doi.org/10.1103/PhysRevD.90.063506}{{\em Phys. Rev.}
  {\bfseries D90} no.~6, (2014) 063506},
\href{http://arxiv.org/abs/1307.8114}{{\ttfamily arXiv:1307.8114 [hep-th]}}.

\bibitem{Weinberg:2003sw}
S.~Weinberg, ``{Adiabatic modes in cosmology},''
  \href{http://dx.doi.org/10.1103/PhysRevD.67.123504}{{\em Phys. Rev.}
  {\bfseries D67} (2003) 123504},
\href{http://arxiv.org/abs/astro-ph/0302326}{{\ttfamily arXiv:astro-ph/0302326
  [astro-ph]}}.

\bibitem{Bassett:2005xm}
B.~A. Bassett, S.~Tsujikawa, and D.~Wands, ``{Inflation dynamics and
  reheating},'' \href{http://dx.doi.org/10.1103/RevModPhys.78.537}{{\em Rev.
  Mod. Phys.} {\bfseries 78} (2006) 537--589},
\href{http://arxiv.org/abs/astro-ph/0507632}{{\ttfamily arXiv:astro-ph/0507632
  [astro-ph]}}.

\bibitem{Senatore:2011sp}
L.~Senatore, E.~Silverstein, and M.~Zaldarriaga, ``{New Sources of
  Gravitational Waves during Inflation},''
  \href{http://dx.doi.org/10.1088/1475-7516/2014/08/016}{{\em JCAP} {\bfseries
  1408} (2014) 016},
\href{http://arxiv.org/abs/1109.0542}{{\ttfamily arXiv:1109.0542 [hep-th]}}.

\bibitem{Mirbabayi:2014jqa}
M.~Mirbabayi, L.~Senatore, E.~Silverstein, and M.~Zaldarriaga, ``{Gravitational
  Waves and the Scale of Inflation},''
  \href{http://dx.doi.org/10.1103/PhysRevD.91.063518}{{\em Phys. Rev.}
  {\bfseries D91} (2015) 063518},
\href{http://arxiv.org/abs/1412.0665}{{\ttfamily arXiv:1412.0665 [hep-th]}}.

\bibitem{Sorbo:2011rz}
L.~Sorbo, ``{Parity violation in the Cosmic Microwave Background from a
  pseudoscalar inflaton},''
  \href{http://dx.doi.org/10.1088/1475-7516/2011/06/003}{{\em JCAP} {\bfseries
  1106} (2011) 003},
\href{http://arxiv.org/abs/1101.1525}{{\ttfamily arXiv:1101.1525
  [astro-ph.CO]}}.

\bibitem{Anber:2012du}
M.~M. Anber and L.~Sorbo, ``{Non-Gaussianities and chiral gravitational waves
  in natural steep inflation},''
  \href{http://dx.doi.org/10.1103/PhysRevD.85.123537}{{\em Phys. Rev.}
  {\bfseries D85} (2012) 123537},
\href{http://arxiv.org/abs/1203.5849}{{\ttfamily arXiv:1203.5849
  [astro-ph.CO]}}.

\bibitem{Dubovsky:2005dw}
S.~L. Dubovsky, P.~G. Tinyakov, and I.~I. Tkachev, ``{Cosmological attractors
  in massive gravity},''
  \href{http://dx.doi.org/10.1103/PhysRevD.72.084011}{{\em Phys. Rev.}
  {\bfseries D72} (2005) 084011},
\href{http://arxiv.org/abs/hep-th/0504067}{{\ttfamily arXiv:hep-th/0504067
  [hep-th]}}.

\end{thebibliography}\endgroup

\end{document}